\documentclass{elsarticle}
\usepackage{tgpagella}
\usepackage[T1]{fontenc}
\usepackage[latin9]{inputenc}
\usepackage{geometry}
\usepackage{verbatim}
\usepackage{amsmath}
\usepackage{amssymb}
\usepackage{graphicx}
\usepackage{esint}

\makeatletter
\usepackage{lineno}
\usepackage{comment}

\newcommand*\patchAmsMathEnvironmentForLineno[1]{%
  \expandafter\let\csname old#1\expandafter\endcsname\csname #1\endcsname
  \expandafter\let\csname oldend#1\expandafter\endcsname\csname end#1\endcsname
  \renewenvironment{#1}%
     {\linenomath\csname old#1\endcsname}%
     {\csname oldend#1\endcsname\endlinenomath}}%
\newcommand*\patchBothAmsMathEnvironmentsForLineno[1]{%
  \patchAmsMathEnvironmentForLineno{#1}%
  \patchAmsMathEnvironmentForLineno{#1*}}%
\AtBeginDocument{%
\patchBothAmsMathEnvironmentsForLineno{equation}%
\patchBothAmsMathEnvironmentsForLineno{align}%
\patchBothAmsMathEnvironmentsForLineno{flalign}%
\patchBothAmsMathEnvironmentsForLineno{alignat}%
\patchBothAmsMathEnvironmentsForLineno{gather}%
\patchBothAmsMathEnvironmentsForLineno{multline}%
}
\usepackage{xcolor}
\usepackage[normalem]{ulem}

\renewcommand{\sout}[1]{}
\newcommand{\green}[1]{#1}
\newcommand{\red}[1]{#1}
\newcommand{\blue}[1]{#1}

\newcommand{\be}[1]{\begin{equation} \label{#1}}
\newcommand{\ee}{\end{equation}}
\newcommand{\bea}[1]{\begin{eqnarray} \label{#1}}
\newcommand{\eea}{\end{eqnarray}}
\newcommand{\difp}[2]{\frac{\partial #1}{\partial #2}}

\biboptions{numbers,sort&compress}
\usepackage[english]{babel}
\addto\captionsenglish{\renewcommand{\appendixname}{Appendix }}

\makeatother

\begin{document}
\global\long\def\tht{\vartheta}%
\global\long\def\ph{\varphi}%
\global\long\def\balpha{\boldsymbol{\alpha}}%
\global\long\def\btheta{\boldsymbol{\theta}}%
\global\long\def\bJ{\boldsymbol{J}}%
\global\long\def\bGamma{\boldsymbol{\Gamma}}%
\global\long\def\bOmega{\boldsymbol{\Omega}}%
\global\long\def\d{\text{d}}%
\global\long\def\t#1{\text{#1}}%
\global\long\def\m{\text{m}}%
\global\long\def\v#1{\boldsymbol{#1}}%
\global\long\def\u#1{\underline{#1}}%

\global\long\def\t#1{\mathbf{#1}}%
\global\long\def\bA{\boldsymbol{A}}%
\global\long\def\bB{\boldsymbol{B}}%
\global\long\def\c{\mathrm{c}}%
\global\long\def\difp#1#2{\frac{\partial#1}{\partial#2}}%
\global\long\def\xset{{\bf x}}%
\global\long\def\zset{{\bf z}}%
\global\long\def\qset{{\bf q}}%
\global\long\def\pset{{\bf p}}%
\global\long\def\wset{{\bf w}}%
\global\long\def\ei{{\bf \mathrm{ei}}}%
\global\long\def\ie{{\bf \mathrm{ie}}}%

\addto\captionsenglish{\renewcommand{\appendixname}{Appendix }}
\hyphenation{gui-ding-cen-ter ca-noni-cal non-ca-noni-cal}

\title{Symplectic integration with non-canonical quadrature for guiding-center
orbits in magnetic confinement devices}

\author{Christopher G. Albert$^{1}$, Sergei V. Kasilov$^{2,3}$, Winfried
Kernbichler$^{2}$}

\address{$^{1}$Max-Planck-Institut für Plasmaphysik, Boltzmannstraße 2, 85748
Garching, Germany\\
$^{2}$Fusion@ÖAW, Institute of Theoretical and Computational Physics,
Graz University of Technology, Petersgasse~16, 8010 Graz, Austria\\
$^{3}$Institute of Plasma Physics, National Science Center ``Kharkov
Institute of Physics and Technology'',\\
Akademicheskaya Str.~1, 61108 Kharkov, Ukraine}
\begin{abstract}
We study symplectic numerical integration of mechanical systems with
a Hamiltonian specified in non-canonical coordinates and its application
to guiding-center motion of charged plasma particles in magnetic confinement
devices. The technique combines time-stepping in canonical coordinates
with quadrature in non-canonical coordinates and is applicable in
systems where a global transformation to canonical coordinates is
known but its inverse is not. A fully implicit class of symplectic
Runge-Kutta schemes has recently been introduced and applied to guiding-center
motion by {[}Zhang et al., Phys. Plasmas 21, 32504 (2014){]}. Here
a generalization of this approach with emphasis on semi-implicit partitioned
schemes is described together with methods to enhance their performance\green{,
in particular avoiding evaluation of non-canonical variables at full
time steps}. For application in toroidal plasma confinement configurations
with nested magnetic flux surfaces a global canonicalization of \blue{coordinates
for} the guiding-center Lagrangian by a spatial transform is presented
that allows for pre-computation of the required map in a parallel
algorithm \blue{in the case of time-independent magnetic field geometry}.
Guiding-center orbits are studied in stationary magnetic equilibrium
fields of an axisymmetric tokamak and a realistic three-dimensional
stellarator configuration. Superior long-term properties of \sout{a
symplectic Euler method} \red{symplectic methods} are demonstrated
in comparison to a conventional adaptive Runge-Kutta scheme. Finally
statistics of fast fusion alpha particle losses over their slowing-down
time are computed in the stellarator field on a representative sample,
reaching a speed-up of the symplectic Euler scheme by more than a
factor three compared to usual Runge-Kutta schemes while keeping the
same statistical accuracy and linear scaling with the number of computing
threads.
\end{abstract}
\date{\today}
\maketitle

\makeatletter 
\def\ps@pprintTitle{ 
 \let\@oddhead\@empty 
 \let\@evenhead\@empty                         \def\@oddfoot{\footnotesize\itshape\hfill\today} \def\@evenfoot{\thepage\hfill}
}
\makeatother

\nolinenumbers

\section{Introduction}

Symplectic integrators \citep{Hairer2002-Geometric,Marsden2001} preserve
the structure of Hamiltonian systems in their discretized form for
numerical treatment. Apart from advantages for theoretical analysis
of such methods a major practical feature of symplectic integrators
is long-term stability at relatively large timesteps and good approximate
conservation of integrals of motion. Symplectic integrators rely on
canonical coordinates which allow for integration schemes that are
easy to implement, ranging from the lowest order explicit-implicit
Euler method to Gauss-Legendre and Lobatto quadrature of arbitrary
high order. Applications of classical symplectic integrators include
long-term tracing of charged particle orbits in electromagnetic fields
of particle accelerators \citep{Forest2006-5321} and in plasmas \citep{Cary1993-98}.
Application to systems with a non-canonical formulation such as motion
of guiding-centers instead of particles, require additional modifications.
Existing methods often rely on explicit expressions to obtain non-canonical
coordinates in terms of canonical ones in simplified geometry \citep{khan2012symplectic,khan2015fast,khan2017symplectic}.
An internal report \citep{cary2018second} not published until recently
already mentions the idea of including transformation equations in
an implicit way that is pursued here. Recently a general way to construct
fully implicit non-canonical symplectic integrators has been applied
using symplectic Runge-Kutta schemes \citep{Zhang2014-32504,Zhu2016}.
The way an integrator is constructed there is based on a construction
of a transformation from non-canonical phase-space coordinates to
canonical ones which is always possible at least locally according
to the Darboux-Lie theorem. More generally such integrators form a
special class of Poisson integrators based on this theorem, described
in Ref.~\citep{Hairer2002-Geometric}, pp.~241-242 where the idea
of the present approach is already outlined but realized only for
explicit inverse transformations from canonical to non-canonical coordinates.
The key observation of Refs.~\citep{cary2018second} and \citep{Zhang2014-32504}
used here is that for a (semi-)implicit integrator it is enough to
know this inverse implicitly. Here we limit ourselves to the case
where global canonical coordinates exist which is realized for the
intended application to guiding-center motion in toroidal fusion devices
with nested magnetic flux surfaces.

Consider a mechanical system with Hamiltonian $H(\zset)$ \blue{and
Poisson brackets $\{\cdot,\cdot\}$} given in non-canonical coordinates
$\zset$ \sout{and} \blue{with} a known \blue{time-independent}
transformation to canonical coordinates $\qset(\zset),\pset(\zset)$.
In a numerical scheme it is possible to solve canonical equations
of motion with non-canonical quadrature points that transform to the
quadrature points from a symplectic method in canonical coordinates.
This involves a root-finding process in order to solve the non-linear
set of coordinate transformation equations in each timestep. It does
however not require the construction of an inverse transformation
$\zset(\qset,\pset)$ and its interpolation in phase-space. As no
further approximations are made, all features of usual symplectic
integrators are retained up to computer accuracy. An interesting property
of such an integrator is that in general the quadrature points for
$\zset$ follow implicitly from the restriction that the method remains
symplectic in the canonical sense in each timestep, which is guaranteed
by fixed quadrature points for $\qset,\pset$. The result of this
approach is a special class of \sout{multistep} multistage methods
with internal stages in non-canonical $\zset$ and with canonical
$\qset,\pset$ kept at full steps. Variants of Gauss-Legendre quadrature
such as the implicit midpoint rule used in \citep{Zhang2014-32504}
are a particular case where non-canonical $\zset$ link $\qset$ and
$\pset$ at the same point in time. Loosening this requirement makes
it possible to formulate semi-implicit \blue{and partitioned} rather
than fully implicit integrators that share their favorable long-term
properties and optionally allow to reconstruct values of $\zset$
at full steps if needed.

In the guiding-center formulation \citep{Littlejohn1983-111,Balescu1988-Transport,Cary2009-693,Possanner2018}
the fast gyration scale of charged particles in a strong magnetic
field is removed by a coordinate transformation in phase-space, effectively
reducing the number of degrees of freedom by one, where the gyrophase
appears as a cyclic variable. Rather than looking at the particle
position, one is interested in the position of the guiding-center,
around which the particle gyrates periodically. In addition to three
coordinates of the guiding-center, the ignorable gyrophase remains
as a fourth position-like coordinate, leaving only two momentum- or
velocity-like coordinates to parametrize six-dimensional phase space.
This imbalance makes it immediately clear that such a description
must be non-canonical. Until recently the only available structure-preserving
integrators for guiding-center motion in general magnetic field geometry
were variational integrators based on discretization of the action
principle with the guiding-center Lagrangian in non-canonical coordinates
\citep{Qin2008-035006,Qin2009-042510}. As this Lagrangian is degenerate,
i.e. doesn't contain kinetic energy as a positive-definite quadratic
form on velocities, variational guiding-center integrators are prone
to \red{multistep parasitic mode} instabilities. This has only recently
been resolved by \sout{either regularization} \blue{using degenerate
variational integrators} or \sout{projection methods} \blue{projected
variational integrators} \citep{Burby2017-110703,Kraus2017-preprint,Morrison2017-55502,Ellison2018}.
Another way to approximately or exactly conserve invariants is possible
via a geometric integrator based on a partition of space into a mesh
\citep{Kasilov2016-282,Eder2019}. Due to low order interpolation,
depending on the symmetry of the mesh, such integrators can introduce
an intolerable amount of artificial stochasticity when used in non-axisymmetric
configurations. Alternative approaches based on symplectic integration
in the classical sense are thus still of interest.

An important result underlying the construction of symplectic integrators
for guiding-center motion is the possibility to construct a transformation
that restores a canonical form of the guiding-center Lagrangian \citep{White1984-2455,Meiss1990-2563,Zhang2014-32504,Li2016-334,Burby2017-110703},
making canonical momenta available as functions of non-canonical phase-space
coordinates. For magnetic configurations with nested flux-surfaces
a purely spatial transformation can be pre-computed globally in a
parallel manner. After that, symplectic integration of an arbitrary
number of orbits in non-canonical coordinates is possible without
additional computational overhead from the canonicalization.

The constructed integrators conserve total energy and parallel adiabatic
invariant within fixed bounds and exactly conserve angular momentum
in axisymmetric configurations. This makes them suitable for tracing
orbits in low-collisional plasma regimes over a long period of time,
including thermal ions in plasma at reactor-relevant conditions as
well as fast fusion $\alpha$ particles especially important for stellarator
optimization~\citep{Lotz1992,Mikhailov2002,Subbotin2006,Drevlak2014b,Nemov2014}.
For practical applications a reduction in computation time compared
to usual Runge-Kutta methods is necessary. In this respect we first
discuss the general form and features of the described integrators
in section~\ref{sec:Implicit-symplectic-integration} with some optimizations
to make them competitive with existing methods in terms of computing
time. Then we proceed to the application to guiding-center motion,
describing canonicalization of coordinates in section~\ref{sec:Canonical-form-of}
and construction of integrators in section~\ref{sec:Integration-of-guiding-center}.
In section~\ref{sec:Results-and-discussion} results are presented
and discussed, and a final summary is given in section~\ref{sec:Summary}.

\section*{Notation}

Here we fix some notational conventions for the following text. Letters
$x^{k}$ are used for spatial coordinates, $q^{i},p_{j}$ for canonical
coordinates and $z^{\alpha}$ for generally non-canonical coordinates
in phase-space. Lower index notation for canonical momenta $p_{j}$
is used since in the classical construction $p_{j}$ are not only
coordinates in phase-space but also components of covectors with respect
to configuration space charted by coordinates $q^{i}$. Latin indices
$i,j,k,l$ run from $1$ to $3$ for spatial coordinates $x^{k}$,
over the number of internal stages if used for an integration scheme,
or from $1$ to $N$ for canonical coordinates in a system with $N$
degrees of freedom. The index $n$ is reserved for notation of time-steps.
Greek indices like $\alpha,\beta$ are used for general phase-space
coordinates $z^{\alpha}$ and run from $1$ to $2N$. We use the usual
convention from tensor algebra to sum over indices appearing once
up and once down in formulas, i.e. $p_{i}\dot{q}^{i}\equiv\sum_{i}p_{i}\dot{q}^{i}$,
where indexes in the denominator of derivatives switch their position,
e.g. $\partial H/\partial q^{j}=p_{j}$ has a lower index. Coordinate
tuples are denoted by bold upright letters $\xset,\zset,\qset,\pset$,
while vectors, covectors and tensors use italic bold letters $\v A,\v B$.
We use the convention to denote functional dependencies such as $\zset(\qset,\pset),\zset(t)$
by the same letters as $\zset$ alone, as often used in physics literature.
If arguments are not written explicitly inside functions the meaning
should become clear from the respective context. Explicit time dependencies
of fields, Lagrangian and Hamiltonian are always suppressed, but the
general results of this work remain valid also in explicitly time-dependent
systems. A notable exception is the transition map from non-canonical
to canonical coordinates that must remain fixed over the integration
time. For guiding-center orbits this limits the application to fixed
or slowly evolving magnetic configurations but permits a time-dependent
electrostatic potential.

\section{Symplectic integrators with non-canonical quadrature points\label{sec:Implicit-symplectic-integration}}

Motivated from the problem of guiding-center motion our goal is to
find symplectic numerical methods for a dynamical system in $2N$-dimensional
phase-space $P$ of a mechanical system with $N$ degrees of freedom
and dynamics specified by a Hamiltonian $H$ \blue{and Poisson brackets
$\{\cdot,\cdot\}$ (or a symplectic structure $\Omega$)}. Phase-points
$z\in P$ are charted in terms of tuples $\zset\equiv(z^{1},z^{2},\dots z^{2N})$
of $2N$ non-canonical coordinates $z^{\alpha}$ via a coordinate
chart map $z(\zset)$. A phase-point $z$ is a coordinate-independent
quantity and not identified with the tuple $\zset$ whose values depend
on the chosen coordinate system if $z$ is fixed. An explicit expression
of the Hamiltonian $H=H(\zset)$ shall exist only in non-canonical
coordinates $\zset$. In addition a transition map i.e. a coordinate
transformation from $\zset$ to $2N$ canonical coordinates $\qset\equiv(q^{1},q^{2},\dots q^{N}),\pset\equiv(p_{1},p_{2},\dots p_{N})$
valid in the phase-space domain of interest shall be given by explicit
and invertible functions
\begin{equation}
q^{i}=q^{i}(\zset),\quad p_{j}=p_{j}(\zset),\label{eq:qk1}
\end{equation}
with inverses $z^{\alpha}(\qset,\pset)$ given only implicitly. 

Equations of motion for the evolution of $q^{i},p_{j}$ can be written
in terms of $\zset$ via
\begin{align}
\frac{\d q^{i}(t)}{\d t} & =\frac{\partial H(\qset(t),\pset(t))}{\partial p_{i}}=\frac{\partial z^{\alpha}(\qset(t),\pset(t))}{\partial p_{i}}\frac{\partial H(\zset(t))}{\partial z^{\alpha}},\label{eq:qdot}\\
\frac{\d p_{j}(t)}{\d t} & =-\frac{\partial H(\qset(t),\pset(t))}{\partial q^{j}}=-\frac{\partial z^{\alpha}(\qset(t),\pset(t))}{\partial q_{j}}\frac{\partial H(\zset(t))}{\partial z^{\alpha}}.\label{eq:pdot}
\end{align}
Derivatives of $z^{\alpha}$ over $q^{i},p_{j}$ are evaluated as
elements of the inverse Jacobian matrix of Eq.~(\ref{eq:qk1}),
\begin{equation}
\frac{\partial\zset}{\partial(\qset,\pset)}=\left(\frac{\partial(\qset,\pset)}{\partial\zset}\right)^{-1},\label{eq:InvJac}
\end{equation}
leaving only dependencies on $\zset(t)$ on the right-hand side of
Eqs.~(\ref{eq:qdot}-\ref{eq:pdot}). Based on an existing symplectic
numerical integration scheme in $\qset,\pset$ discrete equivalents
of the $2N$ equations~(\ref{eq:qdot}-\ref{eq:pdot}) augmented
by $2N$ constraints~(\ref{eq:qk1}) are solved implicitly in $4N$
variables $\qset,\pset,\zset$ via a root-finding method in each timestep.
Quadrature points $q^{i}(t_{i}),p_{j}(t_{j})$ of a chosen symplectic
method for canonical variables fix the respective evaluation point
$z$ in phase-space. Quadrature points of non-canonical variables
$z^{\alpha}$ follow implicitly by the fact that they should describe
the same point $z$ in phase-space via the $2N$ equations~(\ref{eq:qk1})
solved for $q^{i}=q^{i}(t_{i})$ and $p_{j}=p_{j}(t_{j})$ up to machine
accuracy. The resulting virtual $z^{\alpha}$ connect positions and
momenta at possibly different times $t_{i},t_{j}$, but for the same
phase-point $z$. The details of this procedure will be illustrated
in the following section for different numerical schemes. Evaluation
of the Hamiltonian $H$ at a point $z$ in phase-space fixed by a
symplectic scheme in canonical coordinates $q^{i},p_{j}$ guarantees
symplecticity of the integration scheme despite evaluation in terms
of non-canonical coordinates $z^{\alpha}$. An explicit proof of symplecticity
for a particular scheme can easily be stated by using the chain rule
inside a standard proof~\citep{Hairer2002-Geometric,Marsden2001}
to express either the co- or contravariant representation of the canonical
two-form in terms of non-canonical coordinates.

\subsection{Semi-implicit schemes of first order\label{subsec:Semi-implicit-1st}}

\begin{figure}
\begin{centering}
\includegraphics{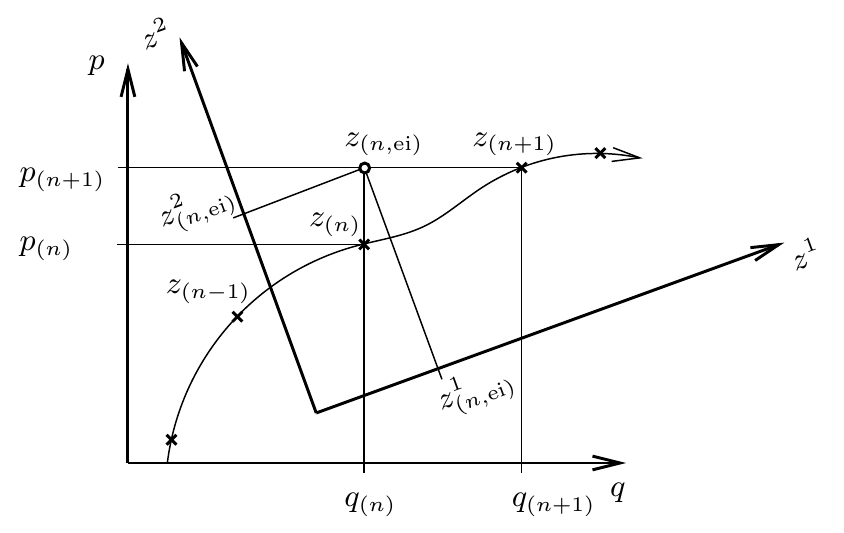}
\par\end{centering}
\caption{Explicit-implicit Euler scheme with coordinate axes for canonical
$(q,p)$ and non-canonical $(z^{1},z^{2})$. The quadrature point
$z_{(n,\protect\ei)}$ plotted as $\circ$ has coordinates $(q_{(n)},p_{(n+1)})$
and $(z_{(n,\protect\ei)}^{1},z_{(n,\protect\ei)}^{2})$. Orbit points
$z_{(n)}=z(t_{(n)})$ plotted as $\times$ have coordinates $(q_{(n)},p_{(n)})$
and $(z_{(n)}^{1},z_{(n)}^{2})$. \label{fig:Explicit-implicit-Euler-scheme}}
\end{figure}

\subsubsection{The explicit-implicit Euler scheme}

Starting at the lowest order we consider the symplectic explicit-implicit
Euler scheme with time-step in canonical coordinates $\qset,\pset$
given by
\begin{align}
p_{j,(n+1)} & =p_{j,(n)}-h\left.\frac{\partial H}{\partial q^{j}}\right|_{\qset_{(n)},\pset_{(n+1)}},\label{eq:imp}\\
q_{(n+1)}^{i} & =q_{(n)}^{i}+h\left.\frac{\partial H}{\partial p_{i}}\right|_{\qset_{(n)},\pset_{(n+1)}}.\label{eq:exp}
\end{align}
Here all position coordinates $\qset_{(n)}\equiv\qset(t_{(n)})$ are
evaluated explicitly at the current time $t_{(n)}$ at the $n$-th
step, and all momentum coordinates $\pset_{(n+1)}\equiv\pset(t_{(n+1)})$
are evaluated implicitly at the subsequent step $t_{(n+1)}\equiv t_{(n)}+h$.
We now modify the scheme in terms of non-canonical quadrature points~$\zset$
that have to describe a phase point $z$ that is charted by canonical
coordinates $\qset_{(n)},\pset_{(n+1)}$ with Eqs.~(\ref{eq:qk1})
being fulfilled. Quadrature points for non-canonical $z^{\alpha}$
with this property are not known a priori, and not necessarily all
$z^{\alpha}$ correspond to coordinate values realized by the actual
orbit at any given time. To reflect this, we use subscript ``ei''
marking the explicit-implicit quadrature scheme to denote the phase-point
$z_{(n,\ei)}$ described by canonical coordinate values $\qset_{(n)},\pset_{(n+1)}$
as well as its non-canonical coordinate values $\zset_{(n,\ei)}$.
The scheme is illustrated in Fig.~\ref{fig:Explicit-implicit-Euler-scheme}
for configuration space dimension $N=1$ .

For the implicit step (\ref{eq:imp}), formally $3N$ equations have
to be solved in variables $\pset_{(n+1)},\zset_{(n,\ei)}$:
\begin{align}
q^{i}(\zset_{(n,\ei)}) & =q_{(n)}^{i},\label{eq:qk}\\
p_{j}(\zset_{(n,\ei)}) & =p_{j,(n+1)},\label{eq:pk}\\
p_{j,(n+1)} & =p_{j,(n)}-h\frac{\partial z_{(n,\ei)}^{\alpha}}{\partial q^{j}}\frac{\partial H}{\partial z^{\alpha}},\label{eq:sympeul1}
\end{align}
where $\partial H/\partial z^{\alpha}$ is evaluated in $\zset=\zset_{(n,\ei)}$
and we have used the short-hand notation
\begin{equation}
\frac{\partial z_{(n,\ei)}^{\alpha}}{\partial q^{j}}\equiv\left.\frac{\partial z^{\alpha}(\qset,\pset)}{\partial q^{j}}\right|_{\zset=\zset_{(n,\ei)}}
\end{equation}
for the pertinent element of the inverse Jacobian matrix~(\ref{eq:InvJac})
of the coordinate transform~(\ref{eq:qk1}). We can eliminate $N$
equations in~(\ref{eq:pk}) right away if we replace $p_{j,(n+1)}$
in Eq.~(\ref{eq:sympeul1}) by the function $p_{j}(\zset_{(n,\ei)})$
and store the result for the subsequent timestep on evaluation. The
final set of $2N$ implicit equation to solve in $\zset_{(n,\ei)}$
is
\begin{align}
q^{i}(\zset_{(n,\ei)})-q_{(n)}^{i} & =0,\label{eq:qimpl0}\\
p_{j}(\zset_{(n,\ei)})-p_{j,(n)}+h\frac{\partial z_{(n,\ei)}^{\alpha}}{\partial q^{j}}\frac{\partial H}{\partial z^{\alpha}} & =0.\label{eq:pimpl1}
\end{align}
Since non-canonical $\zset_{(n,\ei)}$ are uniquely linked to a phase-point
$z_{(n,\ei)}$ with canonical coordinates $\v{\qset}_{(n)},\pset_{(n+1)}$
the problem is well-posed and has a unique solution in $\zset_{(n,\ei)}$
within the stability bounds of $h$ for the underlying symplectic
scheme in $\v{\qset}_{(n)},\pset_{(n+1)}$. After solving Eqs.~(\ref{eq:qimpl0}-\ref{eq:pimpl1}),
new canonical position values $\qset_{(n+1)}\equiv\qset(t_{n+1})$
are explicitly given via (\ref{eq:exp}),
\begin{align}
q_{(n+1)}^{i} & =q_{(n)}^{i}+h\frac{\partial z_{(n,\ei)}^{\alpha}}{\partial p_{i}}\frac{\partial H}{\partial z^{\alpha}}.\label{eq:qn+1}
\end{align}

\subsubsection{The implicit-explicit Euler scheme}

The implicit-explicit Euler scheme is the adjoint to the explicit-implicit
scheme, i.e. the schemes transform into each other under time-reversal.
In the implicit-explicit Euler scheme quadrature points of $\qset$
and $\pset$ are swapped compared to the explicit-implicit Euler scheme
and the implicit system to solve in $\zset_{(n,\ie)}$ (``ie'' for
``implicit-explicit'') becomes
\begin{align}
p_{j}(\zset_{(n,\ie)})-p_{j,(n)} & =0,\label{eq:pimpl0}\\
q^{i}(\zset_{(n,\ie)})-q_{(n)}^{i}-h\frac{\partial z_{(n,\ie)}^{\alpha}}{\partial p_{i}}\frac{\partial H}{\partial z^{\alpha}} & =0.\label{eq:qimpl1}
\end{align}
After finding a sufficiently converged solution to Eqs.~(\ref{eq:pimpl0}-\ref{eq:qimpl1})
we explicitly evaluate
\begin{equation}
p_{j,(n+1)}=p_{j,(n)}-h\frac{\partial z_{(n,\ie)}^{\alpha}}{\partial q^{j}}\frac{\partial H}{\partial z^{\alpha}}.\label{eq:pn+1}
\end{equation}

\subsection{Higher order schemes\label{subsec:Higher-order-schemes}}

\begin{figure}
\begin{centering}
\includegraphics{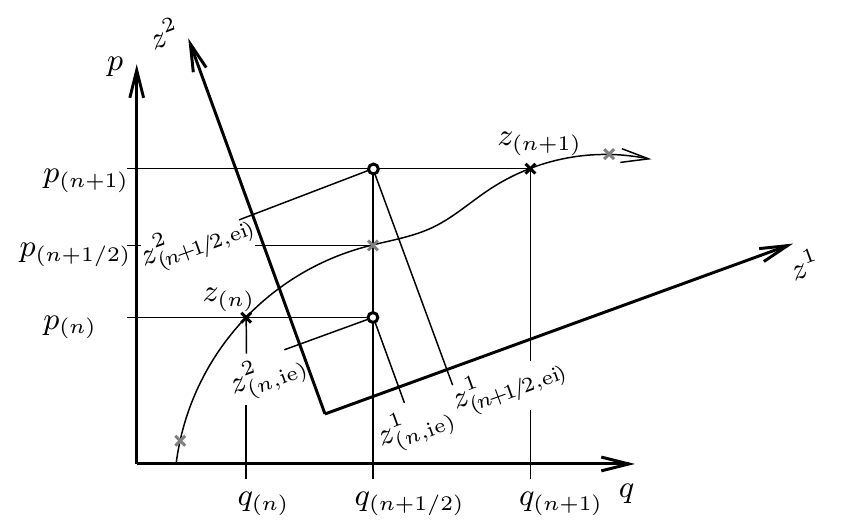}
\par\end{centering}
\caption{Verlet scheme as a combination of two Euler schemes illustrated in
Fig.~\ref{fig:Explicit-implicit-Euler-scheme}. Here also half-step
canonical variables $q_{(n+1/2)},p_{(n+1/2)}$ appear in the symmetric
scheme with non-canonical internal stages $z_{(n,\protect\ie)}^{\alpha},z_{(n+1/2,\protect\ei)}^{\alpha}$
resulting in visibly less bias compared to a single symplectic Euler
step. \label{fig:Explicit-implicit-Euler-scheme-1}}
\end{figure}

Despite their symplectic property, the accuracy of the two partitioned
Euler schemes suffers from their low order and their time-asymmetry,
leading to significant distortion of phase-space when large time-steps
are used \citep{Hairer2002-Geometric}. For applications where this
behavior must be avoided the self-adjoint/time-reversible second-order
\sout{Störmer-}Verlet (leapfrog) algorithm can be used. It is constructed
as a combination of the two partitioned Euler schemes in the following
way (Fig.~\ref{fig:Explicit-implicit-Euler-scheme-1}).
\begin{enumerate}
\item Perform an implicit-explicit Euler step replacing $h$ by $h/2$,
and $(n+1)$ by $(n+1/2)$. The solution of Eqs.~(\ref{eq:pimpl0}-\ref{eq:qimpl1})
yields intermediate non-canonical variables $\zset_{(n,\ie)}$ and
midpoint positions $\qset_{(n+1/2)}$. Midpoint momenta $\pset_{(n+1/2)}$
follow via~(\ref{eq:pn+1}).
\item Perform an explicit-implicit Euler step replacing $h$ by $h/2$,
and $(n)$ by $(n+1/2)$. The solution of Eqs.~(\ref{eq:qimpl0}-\ref{eq:pimpl1})
yields intermediate non-canonical variables $\zset_{(n+1/2,\ei)}$
replacing $\zset_{(n,\ei)}$ with new momenta $\pset_{(n+1)}$ and
finally new positions $\qset_{(n+1)}$ via Eq.~(\ref{eq:qn+1}).
\end{enumerate}

The described semi-implicit symplectic Euler and Verlet methods are
partitioned into implicit and explicit steps. Also the fully implicit
second-order midpoint rule used in \citep{Zhang2014-32504} can be
formulated in the way described above via \blue{non-canonical midpoint
variables $\zset_{(n+1/2)}$ and full step variables $\zset_{(n+1)}$
at the next timestep,}
\begin{align}
q^{i}(\zset_{(n+1/2)}) & =\frac{q_{(n+1)}^{i}+q_{(n)}^{i}}{2},\quad q^{i}(\zset_{(n+1)})=q_{(n+1)}^{i},\label{eq:14-1}\\
p_{j}(\zset_{(n+1/2)}) & =\frac{p_{j\,(n+1)}+p_{j\,(n)}}{2},\quad p_{j}(\zset_{(n+1)})=p_{j\,(n+1)},\label{eq:15}\\
q_{(n+1)}^{i} & =q_{(n)}^{i}+h\frac{\partial z_{(n+1/2)}^{\alpha}}{\partial p_{i}}\frac{\partial H}{\partial z^{\alpha}},\label{eq:16}\\
p_{j,(n+1)} & =p_{j,(n)}-h\frac{\partial z_{(n+1/2)}^{\alpha}}{\partial q_{j}}\frac{\partial H}{\partial z^{\alpha}}.\label{eq:17}
\end{align}
\blue{Here derviatives are evaluated at the internal stage $z_{(n+1/2)}$
with non-canonical coordinates $\zset_{(n+1/2)}$.} By using $q_{(n)}^{i}=q^{i}(\zset_{(n)})$,
$p_{j\,(n)}=p_{j}(\zset_{(n)})$, and eliminating $q_{(n+1)}^{i},p_{j,(n+1)}$
one obtains the form
\begin{align}
q^{i}(\zset_{(n+1/2)}) & =q^{i}(\zset_{(n)})+\frac{h}{2}\frac{\partial z_{(n+1/2)}^{\alpha}}{\partial p_{i}}\frac{\partial H}{\partial z^{\alpha}},\label{eq:14-1-1}\\
p_{j}(\zset_{(n+1/2)}) & =p_{j}(\zset_{(n)})-\frac{h}{2}\frac{\partial z_{(n+1/2)}^{\alpha}}{\partial q_{j}}\frac{\partial H}{\partial z^{\alpha}},\label{eq:15-1}\\
q^{i}(\zset_{(n+1)}) & =q^{i}(\zset_{(n)})+h\frac{\partial z_{(n+1/2)}^{\alpha}}{\partial p_{i}}\frac{\partial H}{\partial z^{\alpha}},\label{eq:znp1}\\
p_{j}(\zset_{(n+1)}) & =p_{j}(\zset_{(n)})-h\frac{\partial z_{(n+1/2)}^{\alpha}}{\partial q_{j}}\frac{\partial H}{\partial z^{\alpha}},\label{eq:znp12}
\end{align}
described in Eq.~(41) of Ref.~\citep{Zhang2014-32504}, where variables
$W$ in the reference correspond to our $\zset_{(n+1/2)}$. It should
be noted that the present treatment is more general than the one of
Ref.~\citep{Zhang2014-32504} in the sense that it doesn't assume
canonical and non-canonical quadrature points to coincide at the same
point in time to first order in each internal stage. Therefore it
is applicable also to partitioned Runge-Kutta methods including the
symplectic Euler and \sout{Störmer-}Verlet schemes described above.
This makes a generalization to higher order schemes via Lobatto \red{IIIA-IIIB}
pairs \red{as well as composition of first order schemes} possible
in addition to Gauss-Legendre quadrature (midpoint for order 2). A
symplectic partitioned Runge-Kutta integrator \citep{Hairer2002-Geometric}
with non-canonical quadrature points $z_{(n,k)}^{\alpha}$ is generally
given by
\begin{align}
q^{i}(\zset_{(n,k)}) & =q_{(n)}^{i}+h\sum_{l}a_{kl}\frac{\partial z_{(n,l)}^{\alpha}}{\partial p_{i}}\frac{\partial H}{\partial z^{\alpha}},\label{eq:rk1}\\
p_{j}(\zset_{(n,k)}) & =p_{j\,(n)}-h\sum_{l}\hat{a}_{kl}\frac{\partial z_{(n,l)}^{\alpha}}{\partial q^{j}}\frac{\partial H}{\partial z^{\alpha}},\\
q_{(n+1)}^{i} & =q_{(n)}^{i}+h\sum_{k}b_{k}\frac{\partial z_{(n,k)}^{\alpha}}{\partial p_{i}}\frac{\partial H}{\partial z^{\alpha}},\label{eq:qnp1}\\
p_{j\,(n+1)} & =p_{j\,(n)}-h\sum_{k}b_{k}\frac{\partial z_{(n,k)}^{\alpha}}{\partial q^{j}}\frac{\partial H}{\partial z^{\alpha}}.\label{eq:rk5}
\end{align}
with $\hat{a}_{kl}=b_{l}-b_{l}a_{lk}/b_{k}$ (no summation over $k$,
$l$) to guarantee symplecticity of the scheme \sout{and Hamiltonian
$H$} evaluated in \red{the phase-point} $z_{(n,k)}$. Here each
component $z_{(n,k)}^{\alpha}$ of \red{non-canonical coordinates}
$\zset_{(n,k)}$ is evaluated independently at a virtual stage such
that Eqs. (\ref{eq:rk1}-\ref{eq:rk5}) are fulfilled, without reference
to time $t$. If only approximate values for $\zset$ are required
it is sufficient to use internal stages $\zset_{(n,k)}$ that are
accurate \sout{within the order of the integration scheme} \red{at
a usually reduced order. This observation is also relevant to non-partitioned
schemes with $\hat{a}_{kl}=a_{kl}$. Also there it is not necessary
to solve full steps in non-canonical variables. The most gain is possible
in the implicit midpoint rule where Eqs.~(\ref{eq:14-1-1}-\ref{eq:15-1})
are replaced by 
\begin{align}
q^{i}(\zset_{(n+1/2)}) & =q_{(n)}^{i}+\frac{h}{2}\frac{\partial z_{(n+1/2)}^{\alpha}}{\partial p_{i}}\frac{\partial H}{\partial z^{\alpha}},\label{eq:14-1-1-1}\\
p_{j}(\zset_{(n+1/2)}) & =p_{j\,(n)}-\frac{h}{2}\frac{\partial z_{(n+1/2)}^{\alpha}}{\partial q_{j}}\frac{\partial H}{\partial z^{\alpha}},\label{eq:15-1-1}
\end{align}
and the remaining equations (\ref{eq:znp1}-\ref{eq:znp12}) corresponding
to Eqs.~(\ref{eq:qnp1}-\ref{eq:rk5}) become explicit with
\begin{align}
q_{(n+1)}^{i} & =q_{(n)}^{i}+h\frac{\partial z_{(n+1/2)}^{\alpha}}{\partial p_{i}}\frac{\partial H}{\partial z^{\alpha}},\label{eq:sisa3}\\
p_{j\,(n+1)} & =p_{j\,(n)}-h\frac{\partial z_{(n+1/2)}^{\alpha}}{\partial q_{j}}\frac{\partial H}{\partial z^{\alpha}}.\label{eq:sisa3-1}
\end{align}
With this modification the dimension of the implicit system to solve
is the same for both first-order Euler variants and the single-stage
but second-order implicit midpoint rule for general non-canonical
coordinates. As we will see in section~\ref{subsec:Semi-implicit-integration-scheme}
the choice of partly canonical variables can reduce the dimension
of symplectic Euler methods, thereby producing truly semi-implicit
schemes.}

In case accurate results for non-canonical variables $\zset_{(n+1)}\equiv\zset(t_{(n+1)})$
are required at full steps, the additional implicit equations
\begin{align}
q^{i}(\zset_{(n+1)}) & =q_{(n+1)}^{i},\\
p_{j}(\zset_{(n+1)}) & =p_{j\,(n+1)}
\end{align}
have to be solved after having obtained $\qset_{(n+1)},\pset_{(n+1)}$
in the time-step, \sout{effectively eliminating the advantage of
semi-implicit schemes in terms of expected} thereby reducing performance.
To avoid this additional overhead one can instead apply a Taylor expansion
\blue{of sufficiently high order} to approximate $\zset_{(n+1)}$
as described in \ref{subsec:Evaluation-of-non-canonical} \blue{in
a post-processing step}.

\newpage{}

\subsection{Numerical solution of implicit equations\label{subsec:Numerical-solution-of}}

\subsubsection{Iterative root-finding schemes and predictor}

For finding roots of implicit sub-steps such as Eq.~(\ref{eq:sympeul1})
usual Newton iterations or similar methods with superconvergence are
well-suited. For the purpose of keeping the method symplectic, convergence
criteria should use tolerance settings close to machine accuracy,
i.e. double-precision floating point values. To retain performance
comparable to explicit methods the number of iterations should be
as low as possible. To achieve this, a good starting guess is required
in order to initialize the root-finding algorithm. We compute an approximate
initial guess for the $k$-th internal stage $z_{(n,k)}^{\alpha}$
via a predictor step. A common way of doing this would consist in
a low-order method such as a forward Euler scheme based on the coordinate
values $z_{(n-1,k)}^{\alpha}$ of the previous full timestep $(n-1)$
via
\begin{align}
z_{(n,k)}^{\alpha} & \approx z_{(n-1,k)}^{\alpha}+\left.h\,\{z^{\alpha},H\}\right|_{\zset=\zset_{(n-1,k)}}.\label{eq:zn+}
\end{align}
Here the time evolution via Poisson brackets,
\begin{align}
\frac{\d z^{\alpha}}{\d t}=\{z^{\alpha},H\} & =\frac{\partial z^{\alpha}}{\partial q^{i}}\frac{\partial H}{\partial p_{i}}-\frac{\partial z^{\alpha}}{\partial p_{i}}\frac{\partial H}{\partial q^{i}}=\left(\frac{\partial z^{\alpha}}{\partial q^{i}}\frac{\partial z^{\beta}}{\partial p_{i}}-\frac{\partial z^{\alpha}}{\partial p_{i}}\frac{\partial z^{\beta}}{\partial q^{i}}\right)\frac{\partial H}{\partial z^{\beta}},\label{eq:poisson}
\end{align}
would be used with derivatives evaluated at $\zset_{(n-1,k)}$ via
the relation of Jacobian matrices in Eq.~(\ref{eq:InvJac}). The
disadvantage of forward predictor steps of this kind and especially
of their higher order equivalents lies in the required extra evaluations
of derivatives of $\qset(\zset),\pset(\zset)$ and $H(\zset)$. An
alternative predictor scheme that has proven to be more efficient
in practice relies on polynomial extrapolation of existing data $z_{(n-1,k)}^{\alpha},z_{(n-2,k)}^{\alpha},\dots$
from multiple previous timesteps. The use of Lagrange polynomials
for this purpose just adds negligible overhead in terms of performance
and storage and doesn't require extra evaluations of fields and derivatives
such as a low-order time-stepping scheme. A starting guess for stages
$\qset_{(n,k)},\pset_{(n,k)}$ can either be obtained in the same
manner, or by inserting the guess for $\zset_{(n,k)}$ into the transformation
(\ref{eq:qk1}) to canonical coordinates.

\subsubsection{Evaluation of fields in the numerical root}

In iterative schemes the fields $\partial z^{\alpha}/\partial q^{i}$
and $\partial H/\partial z^{\alpha}$ are usually not evaluated in
the converged solution, but only in the last iteration step $[m-1]$
before convergence has been reached. For example in the explicit-implicit
Euler scheme we first find the solution $\zset_{(n,\ei)}^{[m]}$ to
(\ref{eq:qimpl0}-\ref{eq:pimpl1}) implicitly in the final iteration,
but have evaluated fields in $\zset_{(n,\ei)}^{[m-1]}$ only in the
previous iteration. In Eq.~(\ref{eq:qn+1}) we however require evaluation
at the converged values $\zset_{(n,\ei)}^{[m]}$. In general, to obtain
field values evaluated at a converged tuple of non-canonical internal
stages $\zset_{(n,k)}^{[m]}$ one can avoid an extra evaluation of
fields using a Taylor expansion around $\zset_{(n,k)}^{[m-1]}$. Namely,
one can express a function $f$ up to first order in differences from
the converged solution as
\begin{equation}
f(\zset_{(n,k)}^{[m]})\approx f(\zset_{(n,k)}^{[m-1]})+(z_{(n,k)}^{\alpha,[m]}-z_{(n,k)}^{\alpha,[m-1]})\frac{\partial f}{\partial z^{\alpha}}\label{eq:taylor}
\end{equation}
with $\partial f/\partial z^{\alpha}$ evaluated at $\zset_{(n,k)}^{[m-1]}$
and summation over $\alpha$. Here $(z_{(n,k)}^{\alpha,[m]}-z_{(n,k)}^{\alpha,[m-1]})$
is of the order of the tolerance setting in the iterative scheme.
For expressions such as $f=\partial H/\partial z^{\alpha}$ we require
second derivatives that can be reused from the Jacobian matrix of
the implicit set of equations solved by the root-finding scheme (not
to be confused with the Jacobian matrix of the transformation to canonical
coordinates). Within a classical (not quasi-) Newton scheme this Jacobian
matrix is available from explicit computations. For practical cases
where just a few iterations are needed, saving one extra evaluation
increases performance considerably.

\blue{Any implicit method can only be symplectic up to the tolerance
in the numerical root finding scheme. With the residual term $(z_{(n,k)}^{\alpha,[m]}-z_{(n,k)}^{\alpha,[m-1]})$
of that order the error amplification in Eq.~(\ref{eq:taylor}) in
the linear term depends on the second factor $\partial f/\partial z^{\alpha}$.
In particular, for guiding-center orbits in an electromagnetic field
this means second spatial derivatives of Hamiltonian $H$ and poloidal
momentum $p_{\vartheta}$, respectively, governed by curvature of
fields. The condition for the validity of the guiding-center approximation
imposes a natural limit for their magnitude. In the practical results
shown below accuracy with and without using Eq.~(\ref{eq:taylor})
is identical at a relative tolerance of $10^{-13}$ for Newton iterations.}

\section{Canonical form of the guiding-center Lagrangian\label{sec:Canonical-form-of}}

The fast gyro-motion of charged particles in a magnetized plasma is
commonly written in terms of the guiding-center Lagrangian~\citep{Littlejohn1983-111,Balescu1988-Transport,Cary2009-693,Possanner2018}
using non-canonical coordinates $\zset=(\xset,\phi,\mu,w)$ in phase
space. Here coordinates $\xset=(r,\tht,\ph)$ parametrize the spatial
position of the guiding-center for which we use the notation with
one radial coordinate $r$ and two angular coordinates $\tht,\ph$
following \citep{Meiss1990-2563}. This describes the important case
of flux coordinates for toroidal fields with embedded closed magnetic
surfaces where a global transformation to canonical coordinates is
possible. In the general case for a transformation constructed only
locally, any coordinate triple may be chosen instead. The pair $(\phi,\mu)$
describes gyrophase and magnetic moment and can be dropped in the
full guiding center Lagrangian, as the guiding-center transformation
is made such that $\phi$ becomes an ignorable variable and $\mu$
becomes a constant of motion. The last coordinate $w$ is often chosen
to be the variable $v_{\parallel}$, corresponding to the velocity
parallel to the magnetic field $\boldsymbol{B}$ within leading order
of the expansion. When leaving $w$ as a general coordinate, $v_{\parallel}=v_{\parallel}(\zset)$
is a function of phase-space coordinates $\zset$. The 4-dimensional
phase-space $P$ relevant for single-particle guiding-center motion
is thus a subspace of the original $6$-dimensional phase-space, containing
points $z$ charted by $\zset=(\xset,w)=(r,\tht,\ph,w)$. It is common
to introduce an effective vector potential $\boldsymbol{A}^{\ast}$
with covariant components
\begin{equation}
A_{k}^{\ast}(\zset)=\frac{mc}{e}v_{\parallel}(\zset)h_{k}(\xset)+A_{k}(\xset),\label{eq:Aeff}
\end{equation}
where $h_{k}$ are the covariant components of the unit vector $\boldsymbol{h}\equiv\v B/B$
along the magnetic field, $\v A$ is the vector potential, and $m,e$
and $c$ are particle mass, charge and speed of light, respectively.
The modified vector potential $\boldsymbol{A}^{\ast}$ depends on
all components of $\zset$, as does the Hamiltonian
\begin{equation}
H(\zset)=\frac{m}{2}v_{\parallel}^{\,2}(\zset)+\mu B(\xset)+e\Phi(\xset).\label{eq:Hz}
\end{equation}
Fields $\v A,\v B,B$ and electrostatic potential $\Phi$ are functions
of the guiding-center position $\xset$, but not of $w$. In principle
quantities can include explicit time dependencies which we do not
specify in the notation. Dropping the ignorable variable $\phi$ and
treating $\mu$ as a constant, the guiding-center Lagrangian describing
the evolution of 4-dimensional phase space coordinates $\zset$ over
time $t$ is given in a non-canonical form,
\begin{equation}
L(\zset,\dot{\zset})=L(\zset,\dot{r},\dot{\tht},\dot{\ph})=\frac{e}{c}\left(A_{r}^{\ast}(\zset)\dot{r}+A_{\tht}^{\ast}(\zset)\dot{\tht}+A_{\ph}^{\ast}(\zset)\dot{\ph}\right)-H(\zset).\label{eq:Lgc1}
\end{equation}
The three terms $\frac{e}{c}A_{i}^{\ast}$ are not canonical momenta
despite their apparent similarity, since only two momenta exist in
a four-dimensional phase space. This asymmetry is a usual property
of the phase space Lagrangian formalism allowing for general, non-canonical
phase space variable transformations which have been employed, in
particular, in~\citep{Littlejohn1983-111}.

\subsection{Transformation to canonical form\label{subsec:Transformation-to-canonical}}

Finding a canonical form of the guiding-center Lagrangian (\ref{eq:Lgc1})
is possible \citep{White1984-2455,Meiss1990-2563,Li2016-334,Burby2017-110703,Albert2016}
while keeping partially non-canonical coordinates. This is usually
achieved by transforming to new phase-space coordinates $\bar{\zset}=(\bar{r},\bar{\tht},\bar{\ph},\bar{w})$
with the constraint that covariant $A_{\bar{r}}^{\ast}$ vanish identically.
The canonical form of the guiding-center Lagrangian $L$ without dependency
on $\dot{\bar{r}}$ follows as
\begin{equation}
L(\bar{\zset},\dot{\bar{\zset}})=L(\bar{\zset},\dot{\bar{\tht}},\dot{\bar{\ph}})=\frac{e}{c}\left(A_{\bar{\tht}}^{\ast}(\bar{\zset})\dot{\bar{\tht}}+A_{\bar{\ph}}^{\ast}(\bar{\zset})\dot{\bar{\ph}}\right)-H(\bar{\zset}).\label{eq:Lgc1-1}
\end{equation}
In this form the coordinates $q^{1}=\bar{\tht}$ and $q^{2}=\bar{\ph}$
appear directly as canonical position coordinates. The remaining part
of the transition map~(\ref{eq:qk1}) from $\bar{\zset}$ to the
two canonical momenta $p_{1}=p_{\bar{\tht}},p_{2}=p_{\bar{\ph}}$
conjugate to $\bar{\tht}$ and $\bar{\ph}$ is directly identified
in the canonical guiding-center Lagrangian (\ref{eq:Lgc1-1}) with
\begin{align}
p_{\bar{\tht}}(\bar{\zset}) & =\frac{e}{c}A_{\bar{\tht}}^{\ast}(\bar{\zset})=mv_{\parallel}(\bar{\zset})h_{\bar{\tht}}(\bar{\xset})+\frac{e}{c}A_{\bar{\tht}}(\bar{\xset}),\label{eq:p}\\
p_{\bar{\ph}}(\bar{\zset}) & =\frac{e}{c}A_{\bar{\ph}}^{\ast}(\bar{\zset})=mv_{\parallel}(\bar{\zset})h_{\bar{\ph}}(\bar{\xset})+\frac{e}{c}A_{\bar{\ph}}(\bar{\xset}).\label{eq:pphi}
\end{align}
Keeping these relations implicit for the construction of a symplectic
scheme in non-canonical coordinates $\bar{\zset}$ as described in
section~\ref{sec:Implicit-symplectic-integration} avoids finding
an inverse that would be required for a usual symplectic integrator
in canonical coordinates.

Canonicalization of phase space coordinates is not unique. It can
be done mixing up spatial guiding center positions with other, ``velocity
space'' variables \citep{White1984-2455,Burby2017-110703,Albert2016,Zhang2014-32504}
or transforming only the spatial coordinates \citep{Meiss1990-2563,Li2016-334}.
In the latter case we can keep $\bar{r}=r$ and denote the angular
variables of canonical coordinates as $\bar{\vartheta}=\vartheta_{\mathrm{c}}$
and $\bar{\varphi}=\varphi_{\mathrm{c}}$. In the next subsection
we construct such a straight field line coordinate system $\xset_{\c}\equiv(r,\vartheta_{c},\varphi_{c})$
similar to the one introduced in Ref.~\citep{Li2016-334} for tokamaks
but with a different method similar to Ref.~\citep{Meiss1990-2563}
which is more efficient, especially for 3D toroidal field configurations
(stellarators) being of interest here. \green{Note that in contrast
to Ref.~\citep{Meiss1990-2563} which results in a non-straight field
line canonical coordinate system the present method produces straight
field line canonical coordinates and requires no special treatment
in case $B_{\vartheta}$ turns to zero which occurs in stellarator
vacuum fields. In case of axisymmetric geometry results are the same
as those of Ref.~\citep{Li2016-334}.}

\subsection{Canonical straight field line flux coordinates for 3D toroidal geometry\label{subsec:Transformation-to-canonical-1}}

Let original coordinates $\xset=(r,\vartheta,\varphi)$ be some general
straight field line magnetic flux coordinates with $B^{r}=0$ and
with vector potential gauge $A_{r}=0$, such that
\begin{equation}
\bA=A_{\vartheta}(r)\nabla\vartheta+A_{\varphi}(r)\nabla\varphi.\label{eq:Aflux}
\end{equation}
These coordinates are linked to any other straight field line flux
coordinates (canonicalized coordinates $\xset_{\c}=(r,\vartheta_{\c},\varphi_{\c})$
in our case) by the transformation (see, e.g., Ref.~\citep{dhaeseleer91})
\begin{equation}
\vartheta=\vartheta_{\c}+\iota(r)G_{\c}(\xset_{\c}),\qquad\varphi=\varphi_{\c}+G_{\c}(\xset_{\c}).\label{eq:invtrans}
\end{equation}
Here, $\iota(r)$ is the rotational transform and $G_{\c}(\xset_{\c})$
is a single valued function (i.e. periodic in angles) which defines
the transformation. Note that it is the inverse coordinate transformation,
$\xset=\xset(\xset_{\c})$, which is constructed here, similarly to
Ref.~\citep{Meiss1990-2563}, in contrast to Ref.~\citep{Li2016-334}
where a direct transformation $\xset_{\c}=\xset_{\c}(\xset)$ has
been looked for. Covariant components of the vector potential transform
by tensor algebra rules and we require to keep the gauge $A_{r}^{\c}\equiv A_{r_{c}}=0$
(here and below various vector components are marked with ``c''
for canonical coordinates). This means simply that the last term containing
$\nabla r$ appearing in direct substitution of Eq.~(\ref{eq:invtrans})
into~(\ref{eq:Aflux}),
\begin{equation}
\bA=A_{\vartheta}(r)\nabla\vartheta_{\c}+A_{\varphi}(r)\nabla\varphi_{\c}+\nabla\left(\iota(r)A_{\vartheta}(r)G_{\c}(\xset_{\c})+A_{\varphi}(r)G_{\c}(\xset_{\c})\right),\label{eq:Aflux_c}
\end{equation}
must be dropped. Here the definition of $\iota(r)$ via derivatives
of the vector potential, $\iota(r)A_{\vartheta}^{\prime}(r)+A_{\varphi}^{\prime}(r)=0$,
has been used. Thus covariant components of $\v A$ corresponding
to magnetic fluxes remain the same also for canonical coordinates,
i.e. $A_{\vartheta}^{\c}=A_{\vartheta}$ and $A_{\varphi}^{\c}=A_{\varphi}$.
An equation for the transformation function $G_{\c}(\xset_{\c})$
leading to $B_{r}^{\c}=0$, which together with $A_{r}^{\c}=0$ annihilates
the radial component $A_{r}^{\ast}$ of the effective vector potential~(\ref{eq:Aeff})
follows from the transformation of the covariant magnetic field components
\begin{equation}
B_{k}^{\c}=B_{j}\difp{x^{j}}{x_{\c}^{k}}.\label{eq:Bcovar}
\end{equation}
Using in the Jacobian matrix $\partial x^{j}/\partial x_{\c}^{k}$,
the explicit form of the coordinate transformation $\xset=\xset(\xset_{\c})$
given by $r=x_{\c}^{1}$ and Eqs.~(\ref{eq:invtrans}) the condition
$B_{r}^{\c}=B_{1}^{\c}=0$ results in the following nonlinear ordinary
differential equation for $G_{\c}$,
\begin{equation}
\difp{}rG_{\c}(\xset_{\c})=-\frac{B_{r}\left(\xset(\xset_{\c})\right)+\iota^{\prime}(r)B_{\vartheta}\left(\xset(\xset_{c})\right)G_{\c}(\xset_{\c})}{\iota(r)B_{\vartheta}\left(\xset(\xset_{\c})\right)+B_{\varphi}\left(\xset(\xset_{\c})\right)}.\label{eq:zeroBr}
\end{equation}
Here and below, for $\xset(\xset_{\c})$ one should use its explicit
form~(\ref{eq:invtrans}) containing $G_{\c}$, and $r=x_{\c}^{1}$.
In turn, the angular components of~(\ref{eq:Bcovar}) define covariant
angular canonical components of the magnetic field via $G_{\c}$ and
its angular derivatives explicitly,
\begin{align}
B_{\vartheta}^{\c}(\xset_{\c}) & =B_{\vartheta}\left(\xset(\xset_{\c})\right)+\left(\iota(r)B_{\vartheta}\left(\xset(\xset_{\c})\right)+B_{\varphi}\left(\xset(\xset_{\c})\right)\right)\difp{}{\vartheta_{\c}}G_{\c}(\xset_{\c}),\nonumber \\
B_{\varphi}^{\c}(\xset_{\c}) & =B_{\varphi}\left(\xset(\xset_{\c})\right)+\left(\iota(r)B_{\vartheta}\left(\xset(\xset_{\c})\right)+B_{\varphi}\left(\xset(\xset_{\c})\right)\right)\difp{}{\varphi_{\c}}G_{\c}(\xset_{\c}).\label{eq:covarBcan}
\end{align}
 The metric determinant of the canonical flux coordinates is also
explicitly defined via $G_{\c}$, its derivatives and the metric determinant
of the original flux coordinates as follows,
\begin{equation}
\sqrt{g_{\c}(\xset_{\c})}=\sqrt{g\left(\xset(\xset_{\c})\right)}\left(1+\iota(r)\difp{G_{c}(\xset_{\c})}{\vartheta_{\c}}+\difp{G_{\c}(\xset_{\c})}{\varphi_{\c}}\right).\label{eq:g_can}
\end{equation}
This yields explicit expressions for the contra-variant field components,
\begin{equation}
B_{\c}^{r}(\xset_{\c})=0,\qquad B_{\c}^{\vartheta}(\xset_{\c})=-\frac{A_{\varphi}^{\prime}(r)}{\sqrt{g_{\c}(\xset_{\c})}},\qquad B_{\c}^{\varphi}(\xset_{\c})=\frac{A_{\vartheta}^{\prime}(r)}{\sqrt{g_{\c}(\xset_{\c})}},
\end{equation}
and, therefore, completes the definition of the magnetic field modulus
$B$ needed in the Hamiltonian~(\ref{eq:Hz}).

Equation~(\ref{eq:zeroBr}) provides a family of solutions which
are defined by the initial condition at the magnetic axis $G_{\c}(0,\vartheta_{\c},\varphi_{\c})=G_{0}(\vartheta_{\c},\varphi_{\c})$.
We use $G_{0}=0$ for well behaved coordinates such that $\varphi_{\c}$
uniquely defines position on the magnetic axis (in a more general
case, $G_{0}$ should be independent of $\vartheta_{\c}$ for well
behaved coordinates).

In our case, coordinates $\xset$ are symmetry flux coordinates~\citep{dhaeseleer91}.
The representation of the magnetic field in these coordinates is a
natural output of the 3D equilibrium code VMEC~\citep{Hirshman1983}.
Using this representation, Eq.~(\ref{eq:zeroBr}) is solved for an
equidistant 2D grid of angles $\vartheta_{\c}$ and $\varphi_{\c}$
which enter as parameters of integration, and the result of this integration
stored on the 3D grid is used for computation of orbits in the form
of 3D double periodic splines. \green{Since Eqs.~(\ref{eq:zeroBr})
are independent for different nodes $(\vartheta_{c},\varphi_{c})$
of the 2D grid in angles, parallelization of canonical coordinate
construction is straightforward.}

\section{Integration of guiding-center orbits\label{sec:Integration-of-guiding-center}}

In the following sections we already assume canonical guiding center
coordinates as constructed in section~\ref{subsec:Transformation-to-canonical}
and omit bars or subscripts ``$\c$'' in the notation of $r,\tht$
and $\ph$. With transformations to canonical coordinates known, guiding-center
orbits can be integrated numerically by the symplectic methods described
in section~\ref{sec:Implicit-symplectic-integration}. In contrast
to Ref.~\citep{Zhang2014-32504} we do not use $v_{\parallel}$ as
a non-canonical coordinate but rather replace it by the canonical
toroidal momentum $p_{\ph}$, leaving the radial coordinate $r$ as
the \emph{only} non-canonical variable in the tuple $\zset\equiv(r,\tht,\ph,p_{\ph})$.
While keeping $r$ is most convenient for the evaluation of magnetic
and electric fields, any quantity $w$ that, together with $\xset=(r,\tht,\ph)$,
uniquely defines a phase-point $z$ can be chosen as a fourth variable,
as mentioned in section~\ref{sec:Canonical-form-of}. Using $w=p_{\ph}$
as the canonical toroidal momentum simplifies expressions for transformations
to canonical coordinates and their derivatives, in particular
\begin{equation}
\frac{\partial r(\tht,\ph,p_{\tht},p_{\ph})}{\partial p_{\tht}}=\left(\frac{\partial p_{\tht}(r,\tht,\ph,p_{\ph})}{\partial r}\right)^{-1}
\end{equation}
as a scalar inverse. For the axisymmetric case in an unperturbed tokamak
field this choice has the further advantage that $p_{\ph}$ is conserved,
allowing for integration in two dimensions with variables $(r,\tht)$.
The definition of $v_{\parallel}$ as a function of $\zset$ is given
by Eq.~(\ref{eq:pphi}) with
\begin{equation}
v_{\parallel}(\zset)=\frac{1}{mh_{\ph}(\xset)}\left(p_{\ph}-\frac{e}{c}A_{\ph}(\xset)\right).\label{eq:vpar}
\end{equation}
The Hamiltonian $H$ follows via~(\ref{eq:Hz}), and the explicit
expression for the poloidal canonical momentum via~(\ref{eq:p}).
First derivatives of the those quantities with respect to spatial
coordinates $\xset$ are
\begin{align}
\frac{\partial v_{\parallel}}{\partial x^{i}} & =-\frac{1}{h_{\ph}}\left(\frac{e}{mc}\frac{\partial A_{\ph}}{\partial x^{i}}+\frac{\partial h_{\ph}}{\partial x^{i}}v_{\parallel}\right),\\
\frac{\partial H}{\partial x^{i}} & =mv_{\parallel}\frac{\partial v_{\parallel}}{\partial x^{i}}+\mu\frac{\partial B}{\partial x^{i}}+e\frac{\partial\Phi}{\partial x^{i}},\\
\frac{\partial p_{\tht}}{\partial x^{i}} & =m\frac{\partial v_{\parallel}}{\partial x^{i}}h_{\tht}+mv_{\parallel}\frac{\partial h_{\tht}}{\partial x^{i}}+\frac{e}{c}\frac{\partial A_{\tht}}{\partial x^{i}}.
\end{align}
Derivatives with respect to $p_{\ph}$ are
\begin{align}
\frac{\partial v_{\parallel}}{\partial p_{\ph}} & =\frac{1}{mh_{\ph}},\quad\frac{\partial H}{\partial p_{\ph}}=mv_{\parallel}\frac{\partial v_{\parallel}}{\partial p_{\ph}}=\frac{v_{\parallel}}{h_{\ph}},\quad\frac{\partial p_{\tht}}{\partial p_{\ph}}=\frac{h_{\tht}}{h_{\ph}}.
\end{align}
Second derivatives required for root-finding methods with user-supplied
Jacobian matrix are given in \ref{subsec:Jacobian-for-root-finding}.

\subsection{Equations of motion for guiding-center orbits\label{subsec:Equations-of-motion}}

Guiding-center dynamics with a \sout{canonicalized} Lagrangian \blue{in
coordinates with $A_{r}^{\star}=0$} are an example of the general
case of ``half-canonical'' coordinates $\zset=(\qset,\wset)$, which
contain canonical positions $q^{i}$ \sout{are} \red{as} a part
of the variable set, but canonical momenta $p_{j}$ are replaced by
non-canonical $w^{k}$. In this case equations of motion (\ref{eq:qdot}-\ref{eq:pdot})
for canonical coordinates, with right-hand side in terms of non-canonical
coordinates are
\begin{align}
\dot{q}^{i} & =\frac{\partial w^{k}}{\partial p_{i}}\frac{\partial H(\qset,\wset)}{\partial w^{k}},\label{eq:qdot-1}\\
\dot{p}_{j} & =-\frac{\partial H(\qset,\wset)}{\partial q^{j}}-\frac{\partial w^{k}}{\partial q^{i}}\frac{\partial H(\qset,\wset)}{\partial w^{k}}.\label{eq:pdot-1}
\end{align}
In the special case of $\qset=(\tht,\ph)$ and $\wset=(r,p_{\ph})$
used for guiding-centers, $r$ is the only remaining non-canonical
phase-space coordinate. Resulting entries of the inverse Jacobian
matrix $\partial\zset/\partial(\qset,\pset)=(\partial(\qset,\pset)/\partial\zset)^{-1}$
of the transition map (\ref{eq:qk1}) to $\qset=(\tht,\ph),\pset=(p_{\tht},p_{\ph})$
are
\begin{align}
\frac{\partial q^{i}(\qset,\pset)}{\partial q^{j}}=\delta_{j}^{i}, & \quad\frac{\partial q^{i}(\qset,\pset)}{\partial p_{\tht}}=0,\quad\frac{\partial q^{i}(\qset,\pset)}{\partial p_{\ph}}=0\\
\frac{\partial r(\qset,\pset)}{\partial q^{j}}=-\left(\frac{\partial p_{\tht}(\zset)}{\partial r}\right)^{-1}\frac{\partial p_{\tht}(\zset)}{\partial q^{j}}, & \quad\frac{\partial r(\qset,\pset)}{\partial p_{\tht}}=\left(\frac{\partial p_{\tht}(\zset)}{\partial r}\right)^{-1},\,\frac{\partial r(\qset,\pset)}{\partial p_{\ph}}=-\left(\frac{\partial p_{\tht}(\zset)}{\partial r}\right)^{-1}\frac{h_{\tht}}{h_{\ph}},\\
\frac{\partial p_{\ph}(\qset,\pset)}{\partial q^{j}}=0, & \quad\frac{\partial p_{\ph}(\qset,\pset)}{\partial p_{\tht}}=0,\quad\frac{\partial p_{\ph}(\qset,\pset)}{\partial p_{\ph}}=1.
\end{align}
Equations of motion (\ref{eq:qdot-1}-\ref{eq:pdot-1}) follow as
\begin{align}
\dot{\tht}(t) & =\frac{\partial H}{\partial r}\left(\frac{\partial p_{\tht}}{\partial r}\right)^{-1},\label{eq:qdot-2}\\
\dot{\ph}(t) & =\frac{\partial H}{\partial p_{\ph}}-\frac{\partial H}{\partial r}\left(\frac{\partial p_{\tht}}{\partial r}\right)^{-1}\frac{\partial p_{\tht}}{\partial p_{\ph}}=\frac{1}{h_{\ph}}\left(v_{\parallel}-\frac{\partial H}{\partial r}\left(\frac{\partial p_{\tht}}{\partial r}\right)^{-1}h_{\tht}\right),\label{eq:phidot}\\
\dot{p}_{\tht}(t) & =-\frac{\partial H}{\partial\tht}+\frac{\partial H}{\partial r}\left(\frac{\partial p_{\tht}}{\partial r}\right)^{-1}\frac{\partial p_{\tht}}{\partial\tht},\label{eq:pdot-2}\\
\dot{p}_{\ph}(t) & =-\frac{\partial H}{\partial\ph}+\frac{\partial H}{\partial r}\left(\frac{\partial p_{\tht}}{\partial r}\right)^{-1}\frac{\partial p_{\tht}}{\partial\ph},
\end{align}
where functions and derivatives on the right-hand side are evaluated
in terms of non-canonical $\zset=(\tht,\ph,r,p_{\ph})$. Finally,
for treatment in general (non-symplectic) integration schemes one
can write the time evolution of non-canonical variables via Poisson
brackets (\ref{eq:poisson}), in particular 
\begin{align}
\dot{r}(t) & =\{r,H\}=-\left(\frac{\partial p_{\tht}}{\partial r}\right)^{-1}\left(\frac{\partial H}{\partial\tht}-\frac{h_{\tht}}{h_{\ph}}\frac{\partial H}{\partial\ph}\right).\label{eq:rdot}
\end{align}

\subsection{Semi-implicit integration schemes for guiding-center motion\label{subsec:Semi-implicit-integration-scheme}}

\subsubsection{Explicit-implicit Euler scheme for guiding-center motion}

For applications where long-term statistical behavior of many orbits
is required rather than absolute accuracy of single orbits it is appropriate
to use the lowest order symplectic Euler schemes. As described in
section~\ref{subsec:Semi-implicit-1st} the first variant of such
an integrator uses quadrature points specified by the symplectic scheme
at fixed time $t_{(n)}$ for \textbf{$(q^{1},q^{2})=(\tht,\ph)$}
and $t_{(n+1)}$ for $(p_{1},p_{2})=(p_{\tht},p_{\ph})$, while the
non-canonical internal stage $r_{(n,\ei)}$ follows implicitly. Three
of the four transformation equations (\ref{eq:qk1}) become trivial.
For the explicit-implicit Euler scheme this means that that $(z_{(n,\ei)}^{2},z_{(n,\ei)}^{3})=(\tht_{(n)},\ph_{(n)})$
are given explicitly. The remaining unknowns $z_{(n,\ei)}^{1}=r_{(n,\ei)}$
and $z_{(n,\ei)}^{4}=p_{\ph,(n+1)}$ appear in a set of two non-linear
implicit equations given by 
\begin{align}
0=F_{1}(r_{(n,\ei)},p_{\ph,(n+1)})\equiv\, & \frac{\partial p_{\tht}}{\partial r}\left(p_{\tht}(r_{(n,\ei)},\tht_{(n)},\ph_{(n)},p_{\ph,(n+1)})-p_{\tht\,(n)}\right)\nonumber \\
 & +h\left(\frac{\partial p_{\tht}}{\partial r}\frac{\partial H}{\partial\tht}-\frac{\partial p_{\tht}}{\partial\tht}\frac{\partial H}{\partial r}\right),\label{eq:FEuler}\\
0=F_{2}(r_{(n,\ei)},p_{\ph,(n+1)})\equiv\, & \frac{\partial p_{\tht}}{\partial r}\left(p_{\ph,(n+1)}-p_{\ph\,(n)}\right)+h\left(\frac{\partial p_{\tht}}{\partial r}\frac{\partial H}{\partial\ph}-\frac{\partial p_{\tht}}{\partial\ph}\frac{\partial H}{\partial r}\right),\label{eq:F2Euler}
\end{align}
with derivatives evaluated at $(r_{(n,\ei)},\tht_{(n)},\ph_{(n)},p_{\ph,(n+1)})$.
Since $\partial p_{\tht}/\partial r$ can in principle vanish at certain
points, we have avoided taking its inverse in the implicit equations.
Angles $\tht_{(n+1)}$ and $\ph_{(n+1)}$ are then computed in an
explicit step by
\begin{align}
\tht_{(n+1)} & =\tht_{(n)}+h\frac{\partial H}{\partial r}\left(\frac{\partial p_{\tht}}{\partial r}\right)^{-1},\label{eq:thnp1}\\
\ph_{(n+1)} & =\ph_{(n)}+h\frac{1}{h_{\ph}}\left(v_{\parallel}-\frac{\partial H}{\partial r}\left(\frac{\partial p_{\tht}}{\partial r}\right)^{-1}h_{\tht}\right),\label{eq:phnp1}
\end{align}
with all functions and derivatives evaluated at $(r_{(n,\ei)},\tht_{(n)},\ph_{(n)},p_{\ph,(n+1)})$
known from the implicit momentum part (\ref{eq:FEuler}-\ref{eq:F2Euler})
of the timestep. In the axisymmetric case where $p_{\ph,(n+1)}=p_{\ph,(n)}=p_{\ph,(0)}$
is conserved, we require only the single equation~(\ref{eq:FEuler})
to be solved in $r_{(n,\ei)}$ for the implicit substep.

\subsubsection{Implicit-explicit Euler scheme for guiding-center motion}

The second possibility for a first-order symplectic integrator is
the implicit-explicit Euler scheme. Here all functions and derivatives
are evaluated in $(r_{(n,\ie)},\tht_{(n+1)},\ph_{(n+1)},p_{\ph,(n)})$.
First we solve the three-dimensional implicit system
\begin{align}
0=F_{1}(r_{(n,\ie)},\tht_{(n+1)},\ph_{(n+1)})\equiv\, & p_{\tht}(r_{(n,\ie)},\tht_{(n+1)},\ph_{(n+1)},p_{\ph,(n)})-p_{\tht,(n)}\label{eq:FEuler-1}\\
0=F_{2}(r_{(n,\ie)},\tht_{(n+1)},\ph_{(n+1)})\equiv\, & \frac{\partial p_{\tht}}{\partial r}(\tht_{(n+1)}-\tht_{(n)})-h\frac{\partial H}{\partial r},\\
0=F_{3}(r_{(n,\ie)},\tht_{(n+1)},\ph_{(n+1)})\equiv\, & h_{\ph}\frac{\partial p_{\tht}}{\partial r}(\ph_{(n+1)}-\ph_{(n)})-h\left(v_{\parallel}\frac{\partial p_{\tht}}{\partial r}-h_{\tht}\frac{\partial H}{\partial r}\right),\label{eq:F2Euler-1}
\end{align}
in the new angles $\tht_{(n+1)},\ph_{(n+1)}$, and $r_{(n,\ie)}$
linking them to $p_{\tht,(n)}$ known from the previous timestep.
New momenta are explicitly computed by
\begin{align}
p_{\tht,(n+1)}=\, & p_{\tht,(n)}-h\left(\frac{\partial H}{\partial\tht}-\frac{\partial H}{\partial r}\frac{\partial p_{\tht}}{\partial\tht}\left(\frac{\partial p_{\tht}}{\partial r}\right)^{-1}\right),\label{eq:FEuler-2}\\
p_{\ph,(n+1)}=\, & p_{\ph,(n)}-h\left(\frac{\partial H}{\partial\ph}-\frac{\partial H}{\partial r}\frac{\partial p_{\tht}}{\partial\ph}\left(\frac{\partial p_{\tht}}{\partial r}\right)^{-1}\right).\label{eq:F2Euler-2}
\end{align}
Higher order methods described in section~\ref{subsec:Higher-order-schemes}
are constructed accordingly.

\section{Results and discussion\label{sec:Results-and-discussion}}

The two variants of symplectic Euler schemes \blue{(first order)},
the Verlet scheme and the implicit midpoint scheme \blue{(second
order)}, \blue{ as well as higher order schemes (Gauss-Legendre,
Lobatto IIIA-IIB pairs, composition methods)} have\sout{ first}
been implemented in \sout{ the programming language Python} \red{Fortran.
Tests are first performed} for the case of an axisymmetric magnetic
field of an idealized tokamak. \sout{The explicit-implicit Euler
scheme has} \red{Both, partitioned and non-partitioned integrators
have} been further optimized by the methods described in section
\ref{subsec:Numerical-solution-of}. \red{The implicit midpoint rule
has been implemented in its original version according to Ref.~\citep{Zhang2014-32504}
and in a single-stage variant according to Eqs.~(\ref{eq:14-1-1-1}-\ref{eq:sisa3-1}).}
The relative performance of different schemes is analyzed and compared
to an adaptive Runge-Kutta 4/5 scheme.\sout{ from the Python library
\emph{SciPy}.} Subsequently \sout{a high-performance version of
the symplectic explicit-implicit Euler scheme of Eqs.~(\ref{eq:FEuler}-\ref{eq:F2Euler})
has been implemented in the Fortran language} \red{the schemes are
used} to compute orbits and loss statistics of fast fusion $\alpha$
particles in a realistic 3D magnetic field of an optimized stellarator~\citep{Drevlak2014}.
There, the preprocessing step described in section~\ref{subsec:Transformation-to-canonical-1}
has been employed to convert magnetic flux coordinates to canonical
form. Results and performance are compared to a Runge-Kutta 4/5 integrator
in usual guiding-center variables using both, VMEC coordinates and
canonical flux coordinates introduced here. This integrator is based
on the guiding-center equations of motion in general curvilinear spatial
coordinates as stated in Ref.~\citep{Nemov2014}.

\red{Evaluation of a 2D magnetic field of a tokamak is computationally
inexpensive, especially when given analytically. Therefore the number
of operations by the integration method itself governs computation
time, e.g. when inverting the Jacobian. In contrast, the cost of 3D
interpolation of a realistic stellarator magnetic field can be higher
than internal computational cost of orbit integration schemes. Therefore
both, CPU time and number of field evaluations are given as independent
performance measures. Accuracy with respect to conservation of invariants
is the main criterion rather than the spatial deviation from a reference
orbit. The reason for this is the intended application to long-term
fusion $\alpha$ particle losses from a 3D volume of a stellarator.
Relative distortion of orbit shapes in phase-space or stretching of
the time coordinate within a few percent is irrelevant for the question
whether an orbit stays confined or not. In contrast, violation of
conservation laws can change orbit topology over time, leading to
unphysical accumulation or ejection of particles in the simulation.
Despite being inherently Hamiltonian, also symplectic methods may
suffer from too high oscillatory deviation of invariants, leading
to artificial stochastic diffusion. The optimum method should require
a minimum amount of CPU time and field evaluations while retaining
invariants within sufficiently small bounds over the physical integration
time.}

\subsection{Guiding-center orbits in an axisymmetric tokamak field}

In axisymmetric geometry the canonicalization of flux coordinates
$(r,\tht,\ph)$ described in section \ref{subsec:Transformation-to-canonical-1}
retains independence from the new toroidal angle $\ph$ for all relevant
quantities, including coordinate transformations, fields and the Hamiltonian.
This permits a formulation with $p_{\ph}=\mathrm{const}$, reducing
the number of variables to just two, $\zset=(r,\tht)$, and results
in integrable motion. Namely, the shape of the orbit is fully determined
by three conservation laws,
\begin{equation}
p_{\ph}=\mathrm{const.},\,\mu=\mathrm{const.},\,H=\mathrm{const.},
\end{equation}
which means that orbits must remain closed in the poloidal projection.
In geometric/symplectic numerical integration schemes this property
is retained~\citep{Hairer2002-Geometric}. The physical time $\tau_{b}$
taken for a single turn until the orbit closes in the poloidal plane
is called a bounce period. 

Here equations of motion are solved numerically by the symplectic
methods described above and compared to an adaptive Runge-Kutta 4/5
(RK) scheme in poloidal variables $(r,\tht)$ based on Eq.~(\ref{eq:qdot-2})
for $\tht$ and Eq.~(\ref{eq:rdot}) for $r$ explained as explained
in section~\ref{subsec:Equations-of-motion}. Those equations can
equivalently be written as
\begin{align}
\frac{\partial p_{\tht}}{\partial r}\dot{\tht}(t) & =\frac{\partial H}{\partial r},\label{eq:eq}\\
\frac{\partial p_{\tht}}{\partial r}\dot{r}(t) & =-\frac{\partial H}{\partial\tht}\underbrace{-\frac{h_{\tht}}{h_{\ph}}\frac{\partial H}{\partial\ph}}_{=0}.\label{eq:hthhph}
\end{align}
Since $\partial H/\partial\varphi$ vanishes due to axisymmetry Eqs.
(\ref{eq:eq}-\ref{eq:hthhph}) take a canonical form in $(\tht,r)$
if time $t$ is replaced by an orbit parameter $\tau$ defined via
\begin{equation}
\d t=\frac{\partial p_{\tht}}{\partial r}\d\tau.\label{eq:dtau}
\end{equation}
Thus for guiding-center dynamics in an axisymmetric magnetic field
formulated in canonicalized flux coordinates one could also use a
usual symplectic scheme in $\tht(\tau),r(\tau)$ with $r$ taking
the role of a canonical momentum, and obtain $t(\tau)$ via (\ref{eq:dtau})
during integration or in a post-processing step. A similar rescaling
of the orbit parameter is used to construct mesh-based geometric integrators
\citep{Kasilov2016-282,Eder2019}.

\subsubsection{Model magnetic field}

For testing purposes we introduce a model magnetic field with $A_{r}=0$
and $B_{r}=0$ from the very beginning, so no canonicalization of
coordinates is required. We use general flux coordinates with flux
surface label $r$ and $B^{r}=0$, which are not necessarily straight
field line coordinates, as e.g. in Ref.~\citep{Meiss1990-2563}.
In this case one is free to choose angular vector potential components
as follows,
\begin{align}
A_{\tht} & =\frac{\partial F}{\partial\tht}+\psi_{\mathrm{tor}}(r),\quad A_{\ph}=\frac{\partial F}{\partial\ph}-\psi_{\mathrm{pol}}(r),\label{eq:athaph}
\end{align}
where $F(r,\tht,\ph)$ is an arbitrary single-valued scalar field,
and $\psi_{\mathrm{tor}}$ and $\psi_{\mathrm{pol}}$ are toroidal
and poloidal magnetic flux normalized by $2\pi$. The remaining quantities
that are free to choose are angular covariant components of $\v h$
and magnetic field modulus $B$. Such an independent choice of fields
does not introduce any contradiction but rather contains the definition
of the metric tensor implicitly. In particular the metric determinant
$\sqrt{g}$ follows via
\begin{align}
B= & h_{\tht}B^{\tht}+h_{\ph}B^{\ph}=h_{\tht}(\nabla\times\v A)^{\tht}+h_{\ph}(\nabla\times\v A)^{\ph}=\frac{1}{\sqrt{g}}\left(h_{\ph}\frac{\partial A_{\tht}}{\partial r}-h_{\tht}\frac{\partial A_{\ph}}{\partial r}\right).
\end{align}
While compatible with Maxwell's equations, the formulation above doesn't
necessarily guarantee a magnetohydrodynamic force balance. For testing
orbit integration the latter is however not relevant. 

Here we choose a setup inspired by the well-known case of the large-aspect
ratio limit in a tokamak with circular, concentric flux surfaces.
The obtained coordinates match quasi-toroidal coordinates up to leading
order in the aspect ratio $r/R_{0}$, where $R_{0}$ is the major
device radius. We define the magnetic field modulus
\begin{align}
B & =B_{0}\left(1-\frac{r}{R_{0}}\cos\tht\right),\label{eq:B}
\end{align}
where the scaling constants $B_{0}$ represents the flux-surface avarage
of $B$. Covariant components of the vector potential are defined
based on Eq.~(\ref{eq:athaph}) with
\begin{align}
A_{r}=0,\quad A_{\tht} & =B_{0}\left(\frac{r^{2}}{2}-\frac{r^{3}}{3R_{0}}\cos\tht\right),\quad A_{\ph}=-\iota_{0}B_{0}\left(\frac{r^{2}}{2}-\frac{r^{4}}{4a^{2}}\right).\label{eq:a}
\end{align}
where $a=\mathrm{const}$ is of the order of the plasma minor radius,
and the rotational transform
\begin{equation}
\iota(r)\equiv\frac{\psi_{\mathrm{pol}}^{\prime}(r)}{\psi_{\mathrm{tor}}^{\prime}(r)}\equiv\iota_{0}\left(1-\frac{r^{2}}{a^{2}}\right)
\end{equation}
has been set to to a linear function of the toroidal flux $\psi_{\mathrm{tor}}$.
In addition we specify
\begin{align}
h_{r}=0,\quad h_{\tht} & =\iota(r)\frac{r^{2}}{R_{0}},\quad h_{\ph}=R_{0}+r\cos\tht.\label{eq:h}
\end{align}
For the metric determinant we obtain
\begin{align}
\sqrt{g} & =r\left(R_{0}+r\cos\vartheta\right)+\frac{\iota^{2}(r)\,r^{3}}{R_{0}-r\cos\vartheta}
\end{align}
which matches $\sqrt{g}$ of geometrical quasi-toroidal coordinates
to linear order in $r/R_{0}$. We thus introduce quasi-cylindrical
coordinates with
\begin{align}
R & =R_{0}+r\cos\tht,\quad Z=r\sin\tht,\label{eq:cyl}
\end{align}
which match usual cylindrical coordinates in the large aspect ratio
limit. For plotting purposes, also in Fig.~\ref{fig:Poloidal-projection}
for the stellarator field, we keep coordinates $(R,Z)$ defined via
Eq.~(\ref{eq:cyl}) valid for any set of flux coordinates $(r,\tht,\ph)$.
If $\sqrt{g}$ is not close to $rR$, such coordinates retain their
topological properties, despite not being close to geometrical cylindrical
coordinates anymore.

\subsubsection{Numerical results of orbits in the axisymmetric model field}

\begin{figure}
\begin{centering}
\includegraphics[scale=0.9]{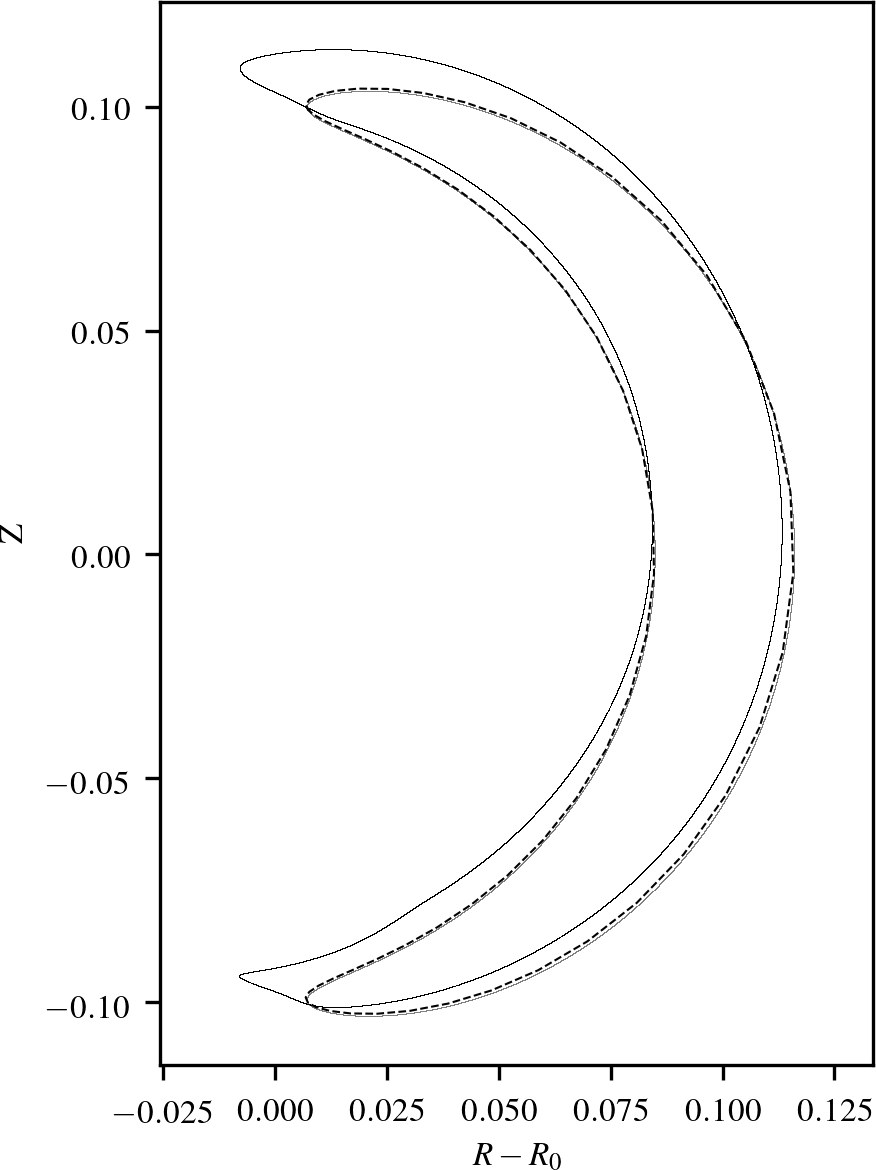}\includegraphics[scale=0.9]{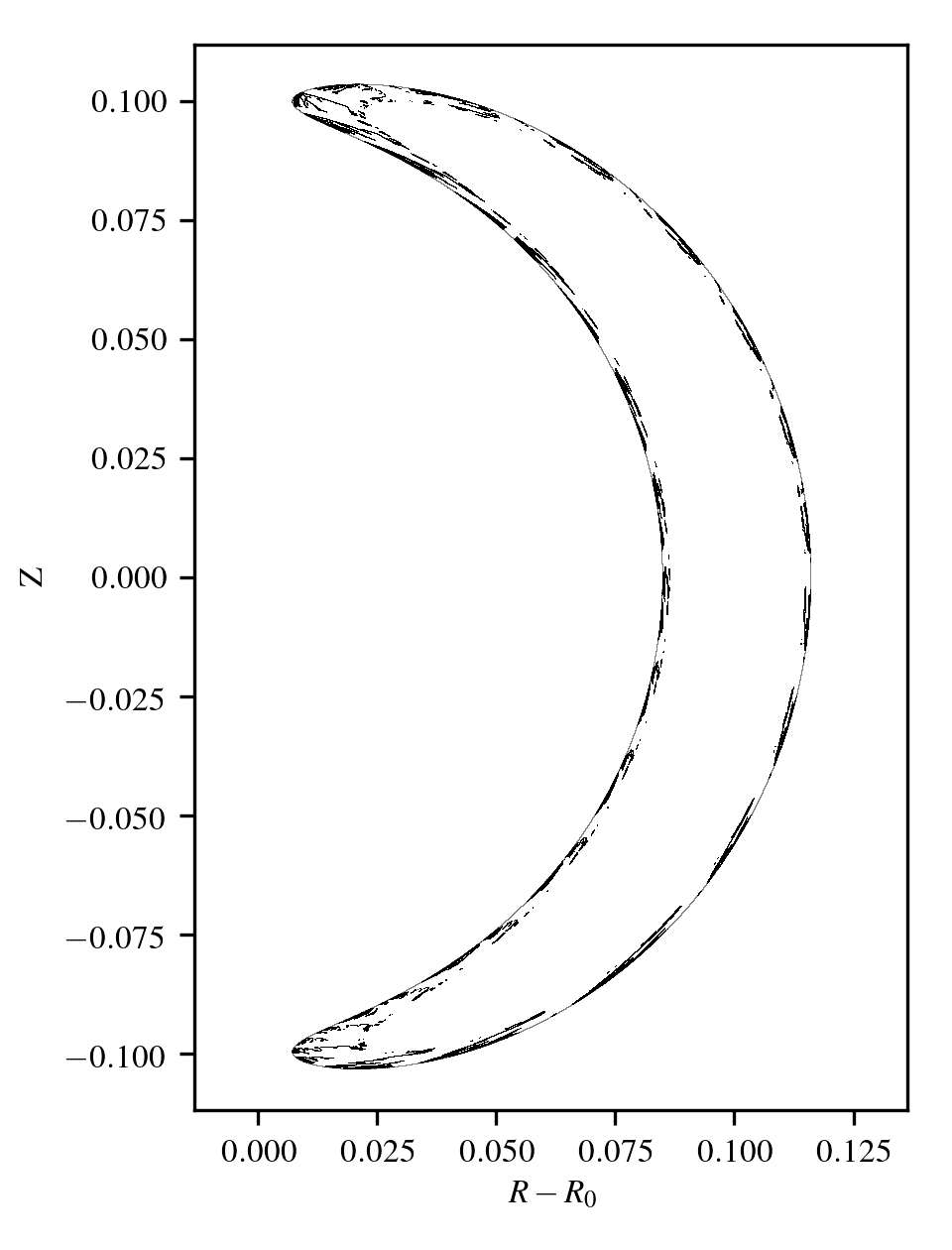}
\par\end{centering}
\caption{Banana orbit in the model field configuration with comparison to reference
(gray). \green{Left:} exlicit-implicit Euler, 8 steps \green{(solid) / 64 steps (dashed)}
per bounce period\sout{ (left)}. \green{Right:} RK4/5 , relative
tolerance of $10^{-6}$\sout{ (right)}.\label{fig:Banana-orbit-for-1}}
\end{figure}

\begin{figure}
\begin{centering}
\includegraphics[scale=0.9]{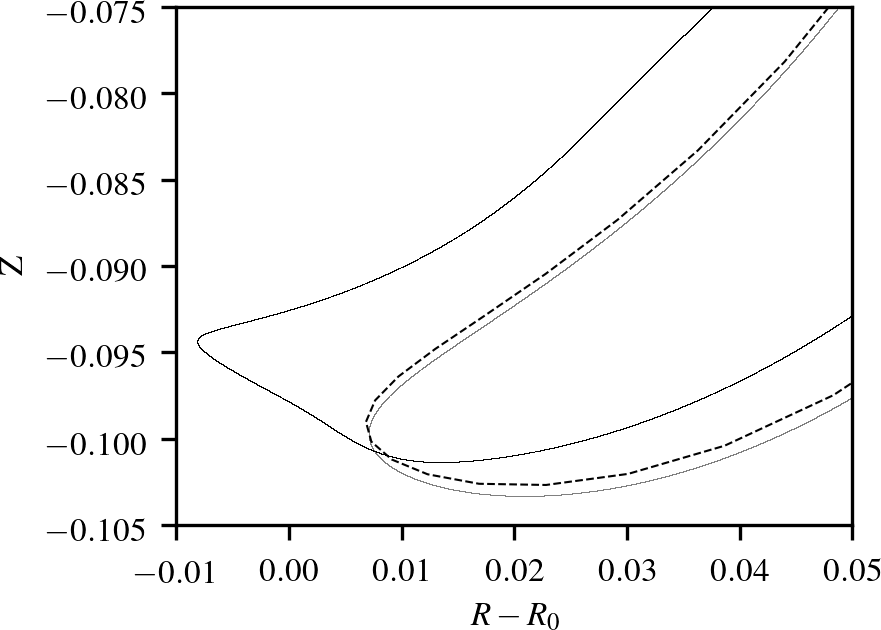}\includegraphics[scale=0.9]{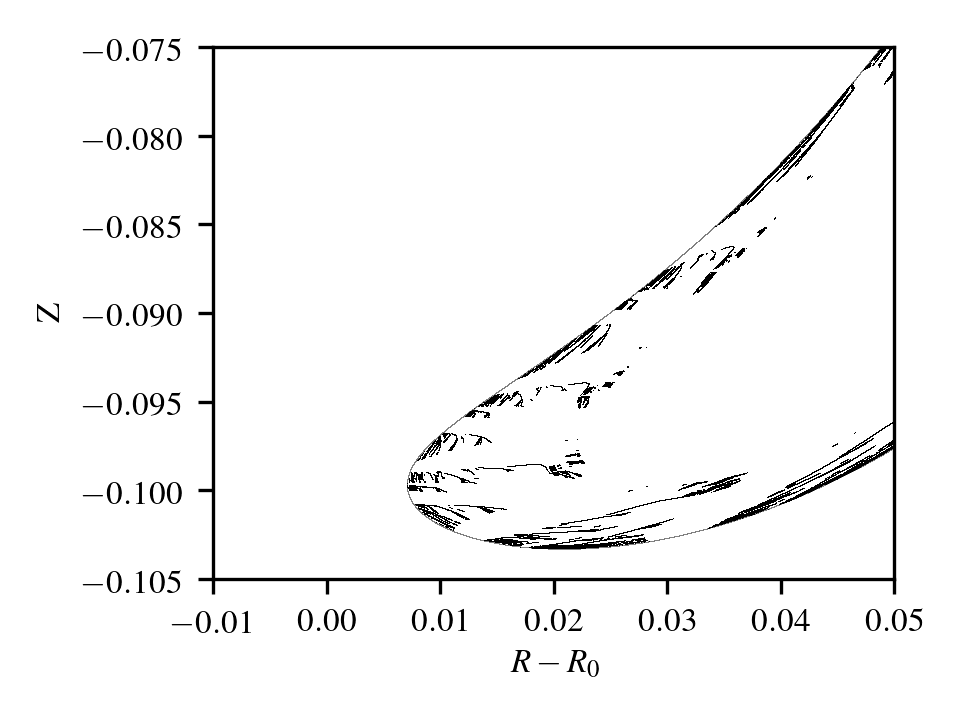}
\par\end{centering}
\caption{Detailed zoom into lower banana tip of Fig.~\ref{fig:Banana-orbit-for-1}.
The symplectic explicit-implicit Euler scheme (left) produces a continuous
closed orbit, while the RK4/5 scheme (right) shows numerical drift
towards the inside of the banana. \label{fig:Banana-orbit-for}}
\end{figure}

Numerical results for a trapped (banana) guiding-center orbit in the
given axisymmetric model field are plotted in Figs.~\ref{fig:Banana-orbit-for-1}-\ref{fig:Banana-orbit-for}
with disconnected points for each timestep. As a reference a single
bounce period was traced with the RK scheme with tolerance at machine
floating-point accuracy. The same orbit was then traced with different
symplectic integrators at low resolution close to their stability
boundary for $\approx10^{5}$ bounce periods. In the symplectic explicit-implicit
Euler scheme using all optimizations described in section~\ref{subsec:Numerical-solution-of}
this corresponded to $\approx8$ steps per bounce period, with $\approx4$
field evaluations per time-step, or $3.2$ million evaluations in
total. As a comparison, a run of the non-symplectic RK4/5 scheme at
the highest possible relative tolerance of $10^{-6}$ that leaves
the distance to the reference orbit well below the banana width after
$10^{5}$ bounce periods, still required more than seven times the
number of field evaluations. \blue{According to the results below
this corresponds to requiring a minimum accuracy is conservation of
invariants within $5\%$ after 3000 bounce periods.} As visible in
Fig.~\ref{fig:Banana-orbit-for-1}, at the majority of points the
RK scheme stays geometrically closer to the reference orbit than the
symplectic Euler scheme. An up-down bias in the plot of the explicit-implicit
Euler orbit is the result of the time-asymmetry of the scheme. While
the orbit points of the symplectic Euler scheme are distributed densely
over a curve in the poloidal projection, orbit points of the RK scheme
are scattered. In particular, at the zoomed picture of the lower banana
tip in Fig.~\ref{fig:Banana-orbit-for} one can see that here the
RK orbit always stays inside the reference orbit, while the symplectic
Euler scheme rather distorts the orbit's shape.
\begin{figure}
\begin{centering}
\includegraphics[scale=0.9]{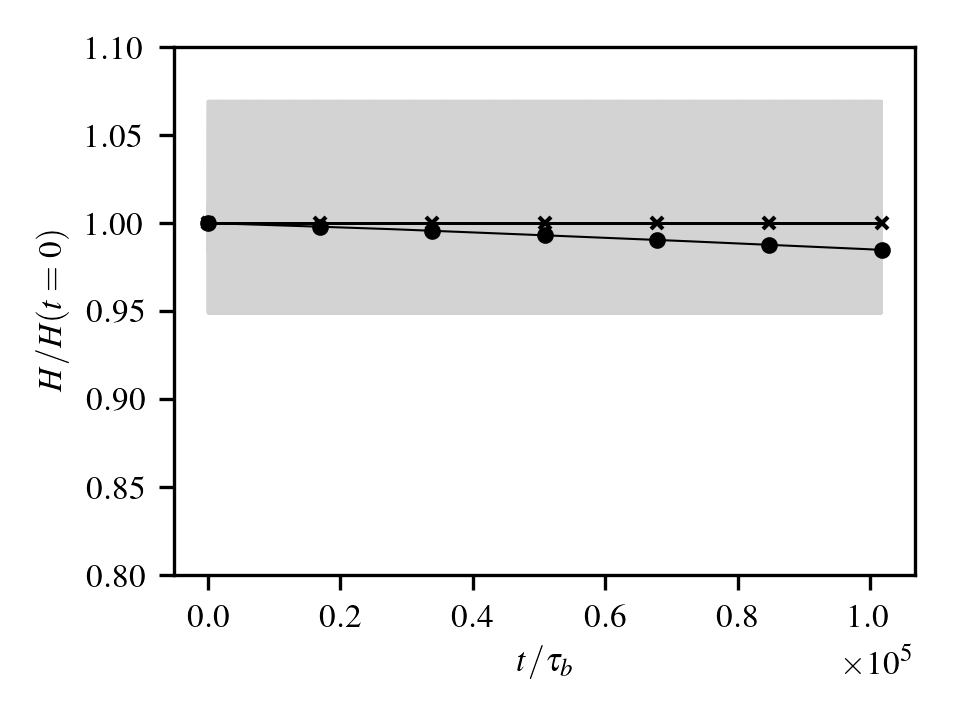}\includegraphics[scale=0.9]{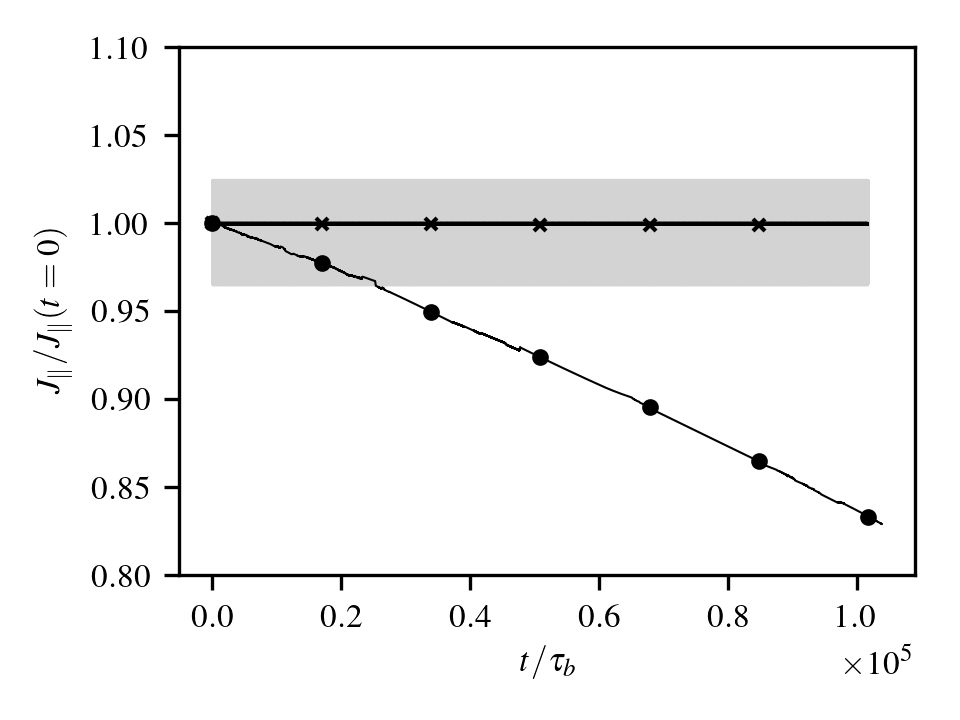}
\par\end{centering}
\caption{Normalized energy (Hamiltonian $H$) an parallel invariant $J_{\parallel}$
over $10^{5}$ bounce periods with integration via RK4/5 ($\bullet$)
compared to the explicit-implicit Euler scheme ($\boldsymbol{\mathbf{\times}}$).
Averaging is performed over $100$ bounce periods. Shaded regions
contain the periodic oscillations of the Euler scheme. \label{fig:Normalized-energy-over}}
\end{figure}

Fig.~\ref{fig:Normalized-energy-over} shows the time evolution of
the Hamiltonian $H$ representing the total energy, and the parallel
adiabatic invariant $J_{\parallel}$. The latter is defined by an
integral over the distance $l$ passed along the field line during
a single bounce period $\tau_{b}$ by a trapped particle as follows,
\begin{equation}
J_{\parallel}\equiv m\oint v_{\parallel}\,\d l=m\int_{0}^{\tau_{b}}v_{\parallel}^{\,2}\,\d t.\label{eq:Jpar}
\end{equation}
In the exact solution both, $H$ and $J_{\parallel}$ are conserved
quantities. The energy of the explicit-implicit Euler integrator oscillates
over a range of several percent, represented by shaded regions in
the plot, while the RK4/5 scheme steadily reduces the orbit energy
over time. Taking an average over $100$ bounce periods however shows
the excellent long-term stability of the symplectic scheme without
any average drift. For the parallel invariant $J_{\parallel}$ the
discrepancy is even more pronounced, with the RK scheme losing about
15\% of the initial value, explaining the inwards drift of the orbit
in Figs.~\ref{fig:Banana-orbit-for-1}-\ref{fig:Normalized-energy-over}.

\begin{figure}
\begin{centering}
\includegraphics[width=0.5\textwidth]{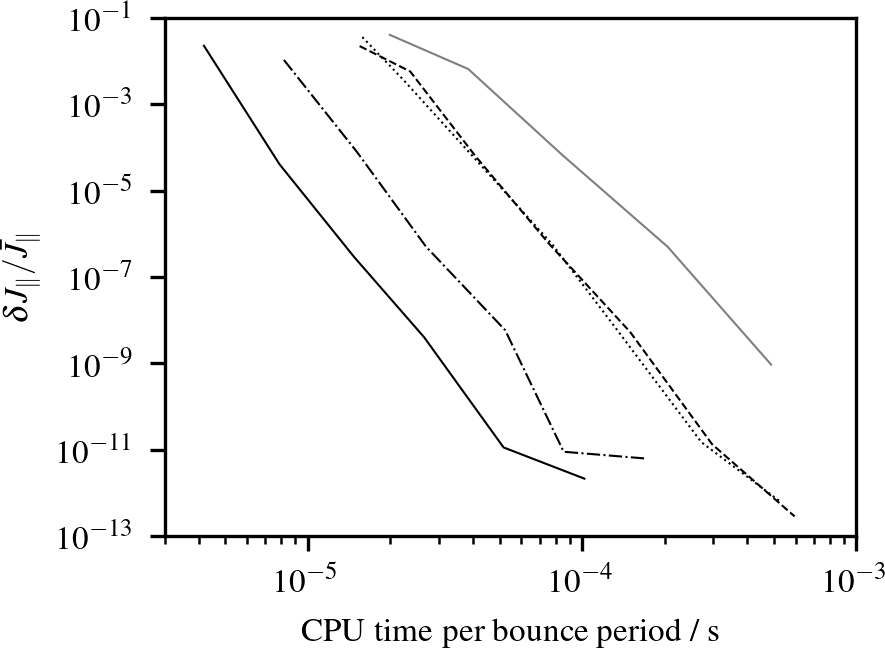}\includegraphics[width=0.5\textwidth]{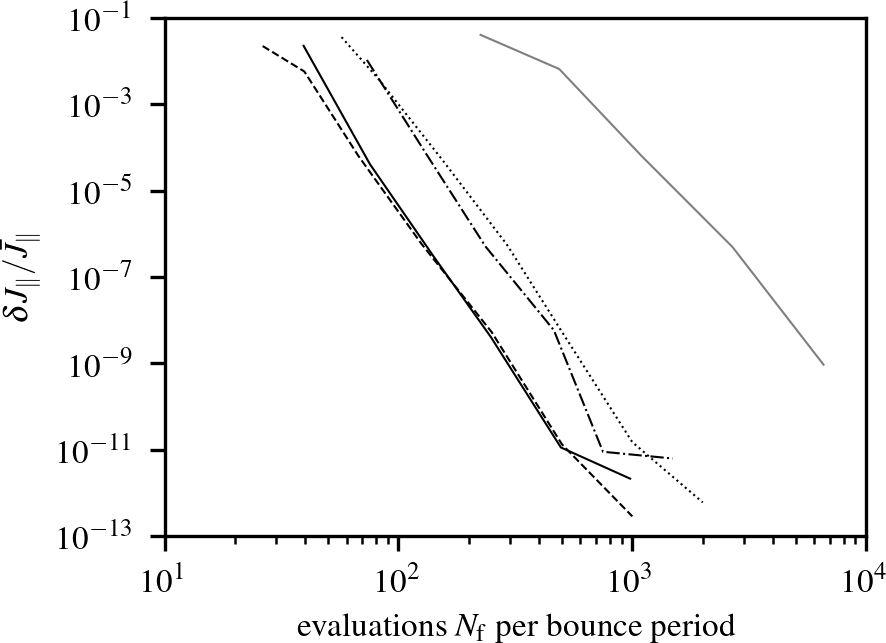}
\par\end{centering}
\caption{\red{Normalized root-mean-square deviation of $J_{\parallel}$ at
different timestep resolutions for a guiding-center orbit in an axisymmetric
tokamak field traced over $10^{5}$ bounce periods. RK4/5 (solid gray),
Explicit-implicit Euler (solid black), implicit midpoint (dotted),
single-stage midpoint (dashed) and Verlet (dash-dotted).} \label{fig:Jpartok}}
\end{figure}

Fig.~\ref{fig:Jpartok} shows a comparison of \red{computation time as well as}
the \sout{minimum} number of field evaluations $N_{\mathrm{f}}$
per bounce period that could be reached in the test case in Figs.~\ref{fig:Banana-orbit-for-1}-\ref{fig:Normalized-energy-over}
\red{for an according accuracy in $J_{\parallel}$ during $N_{\mathrm{b}}=10^{5}$
bounce periods. As an accuracy measure the root-mean-square deviation
$\delta J_{\parallel}$ over all, $N_{b}$, bounce periods is used
with
\begin{equation}
\delta J_{\parallel}=\sqrt{\frac{1}{N_{b}}\sum_{n=1}^{N_{\mathrm{b}}}\left(J_{\parallel}(n\tau_{b})-\bar{J}_{\parallel}\right)^{2}},\qquad\bar{J}_{\parallel}=\frac{1}{N_{b}}\sum_{n=1}^{N_{\mathrm{b}}}J_{\parallel}(n\tau_{b}).
\end{equation}
Results are shown} for different first and second order symplectic
schemes compared to the adaptive Runge-Kutta 4/5 scheme \sout{implemented
in the routine \emph{RK45} within \emph{SciPy} } according to Ref.~\citep{Dormand1980}.
\sout{The explicit-implicit scheme has been implemented in an optimized
variant using} \red{ Implicit systems are solved via} Newton iterations
with a user-supplied Jacobian matrix for root-finding and extrapolating
fields to the final iteration as described in section~\ref{subsec:Numerical-solution-of}.
\sout{The remaining cases use a generic root-finding scheme with
approximated Jacobian implemented in the \emph{fsolve} routine of
\emph{SciPy} that is based on the Powell hybrid method of \emph{MINPACK}.
In addition, the equivalent number of first-order time steps per bounce
period is provided. This number is equal to the number of steps times
the number of internal stages, which are equivalent to a single Euler
step, i.e. the step count times two for Verlet and midpoint rule.}
In the test case all symplectic methods outperform the RK4/5 scheme
and the \red{root-mean-square} deviation \red{$\delta J_{\parallel}$
from its mean value rapidly scales with order $N_{\mathrm{f}}^{-K}$
down to computer accuracy, with $K\approx10$ independently from the
method}. The number of field evaluations $N_{\mathrm{f}}$ per bounce
period \sout{, being the main criterion,} is lowest for the explicit-implicit
Euler scheme and the single stage implicit midpoint scheme \red{
with similar performance of the implicit-explicit Euler scheme not
shown in the graph. Computation time is lowest for the explicit-implicit
Euler scheme, producing only a $1D$ implicit system in $r$ here.
This simplicity favors also the partitioned Verlet scheme over the
implicit midpoint rule in terms of CPU time in the test case.} \sout{and
can be further reduced by the described optimizations. Relatively
poor performance of the Verlet scheme can be observed, reaching its
stability boundary already at half the sub-step width compared to
using either of the two symplectic Euler schemes or the midpoint rule.}

\begin{figure}
\begin{centering}
\includegraphics[width=0.5\textwidth]{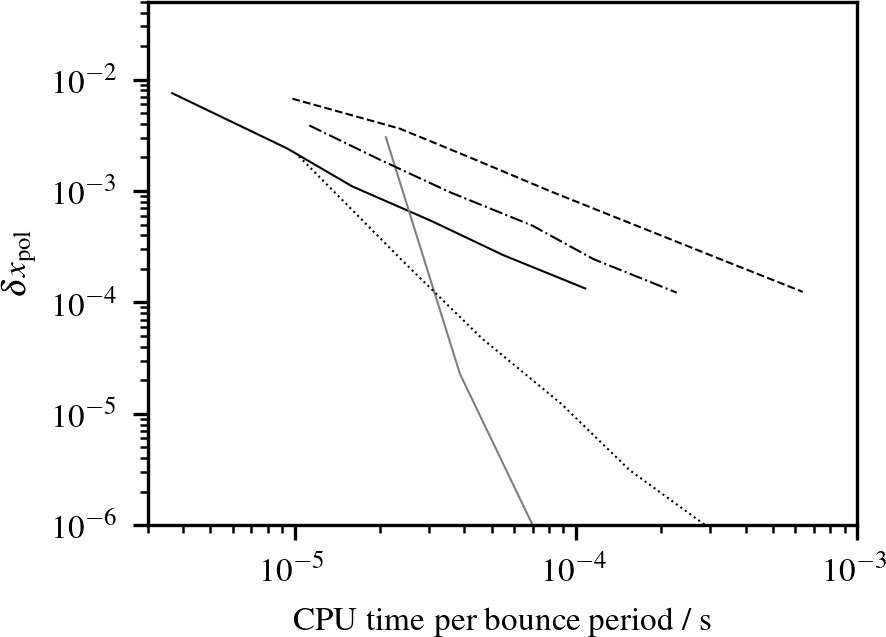}\includegraphics[width=0.5\textwidth]{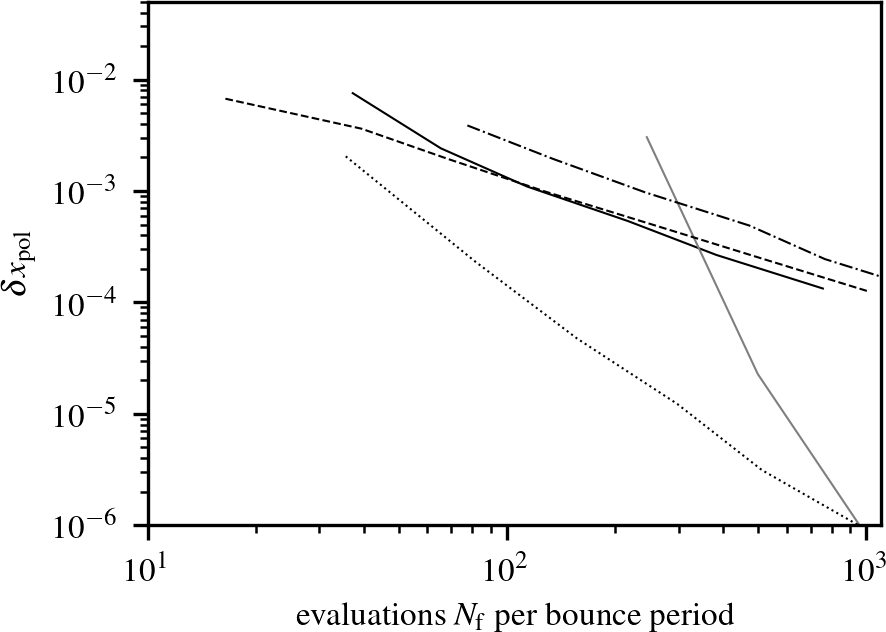}
\par\end{centering}
\caption{\red{Root-mean-square distance to reference orbit in the poloidal
$r-\vartheta$ projection over $10^{5}$ bounce periods. RK4/5 (solid
gray), Explicit-implicit Euler (solid black), implicit midpoint (dotted).
Single-stage midpoint (dashed) and Verlet (dash-dotted) show only
first order accuracy due to direct use of internal stages to save
transformations at full steps.} \label{fig:distance1}}
\end{figure}

Fig.~\ref{fig:distance1} shows the scaling of the geometrical distance
\red{(root-mean-square over integration time of distance to nearest
reference orbit point in each step)} from the reference orbit with
$N_{\mathrm{f}}$ via different step sizes. The superior scaling in
terms of \sout{spatial} \blue{this} accuracy of the implicit midpoint
rule compared to the explicit-implicit Euler method is apparent, yielding
scaling of order $N_{\mathrm{f}}^{-2.6}$ in this example, compared
to order $N_{\mathrm{f}}^{-1.3}$ of the Euler scheme. \red{In contrast
to case of the invariant $J_{\parallel}$ the scaling of the spatial
distance corresponds to the usual order of such integrators. The RK4/5
integrator becomes most accurate at a certain point. The single-stage
midpoint rule and the Verlet method retain only first-order accuracy
due to the uncorrected use of points from internal stages. In terms
of field evaluations per bounce period the single-stage midpoint rule
reaches an impressive minimum of $N_{\mathrm{f}}=16$. Explicit-implicit
Euler and original midpoint rule require about twice as many evaluations.
These values are for tolerance of $10^{-13}$ in the implicit system
being required for long-term stability. } \sout{If however the absolute
number of field evaluations is a criterion rather the high spatial
accuracy, the optimized explicit-implicit Euler scheme can be realized
with a minimum $N_{\mathrm{f}}=32$ evaluations per bounce period
of the considered orbit compared to $N_{\mathrm{f}}=133$ evaluations
for the implicit midpoint rule.}

\subsection{Guiding-center orbits in a three-dimensional stellarator field\label{subsec:Guiding-center-orbits-in}}

In this section results of symplectic orbit integration and the comparison
to currently used methods are presented for a 3D stellarator field
configuration described in Ref.~\citep{Drevlak2014}. This is a quasi-isodynamic
reactor-scale device with five field periods and major radius of $25\,\mathrm{m}$.
The magnetic field has been normalized here so that its modulus averaged
over Boozer coordinate angles on the starting surface is $B_{00}=5\,\mathrm{T}$.
We trace $3.5\,\mathrm{MeV}$ fusion $\alpha$ particles from $t=0$
until $t=1\,\mathrm{s}$, which is of the order of their slowing-down
time in the reactor. As the goal is to reduce the number of field
evaluations as far as possible while retaining physically meaningful
but not necessarily geometrically accurate orbits, \red{low-order
methods will be shown to be best-suited for this task.}\sout{the
optimized variant of the explicit-implicit Euler scheme is chosen
rather than a higher-order method.}

\subsubsection{Numerical results for a single orbit}

\begin{figure}
\begin{centering}
\includegraphics[scale=0.9]{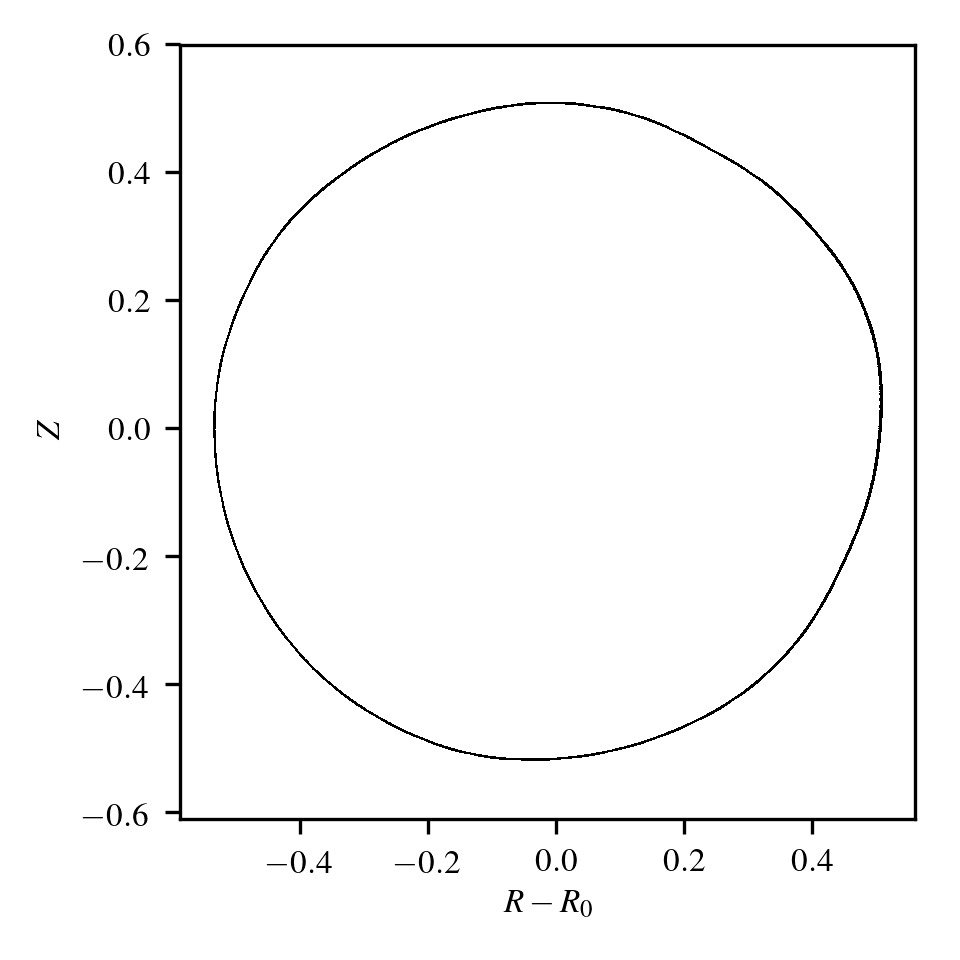}\includegraphics[scale=0.9]{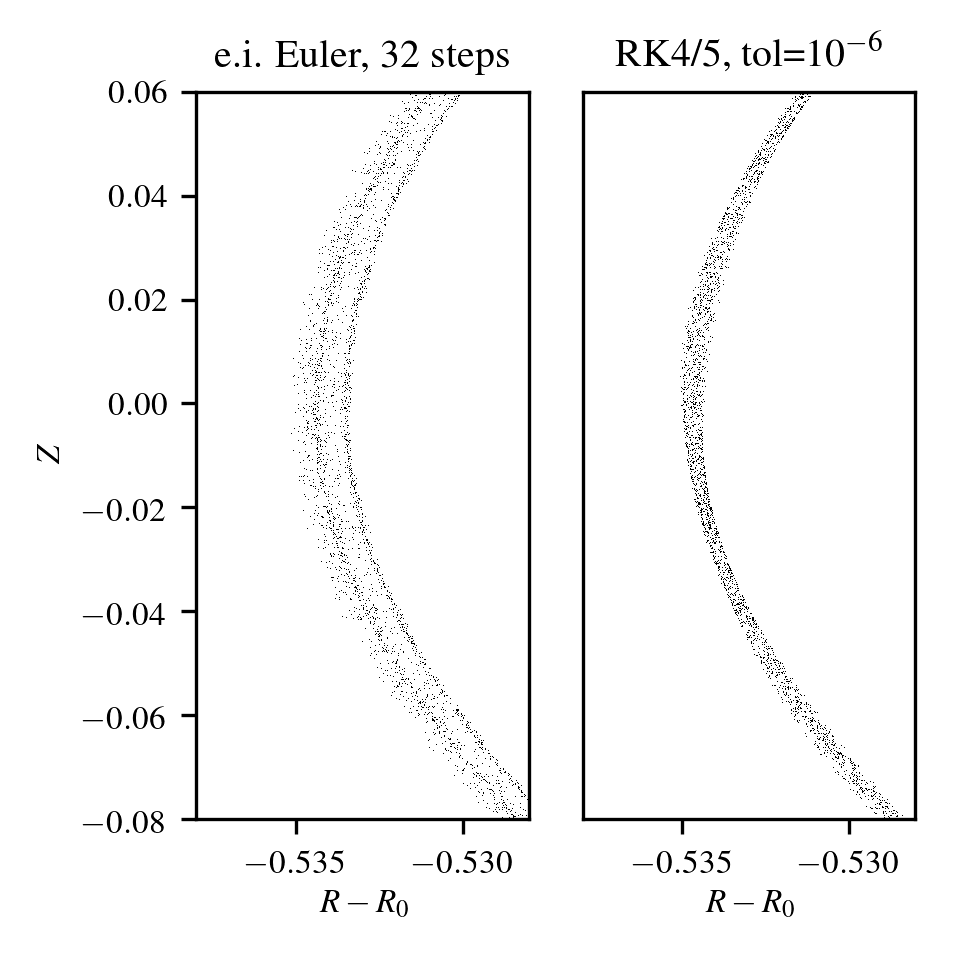}\\
\includegraphics[scale=0.9]{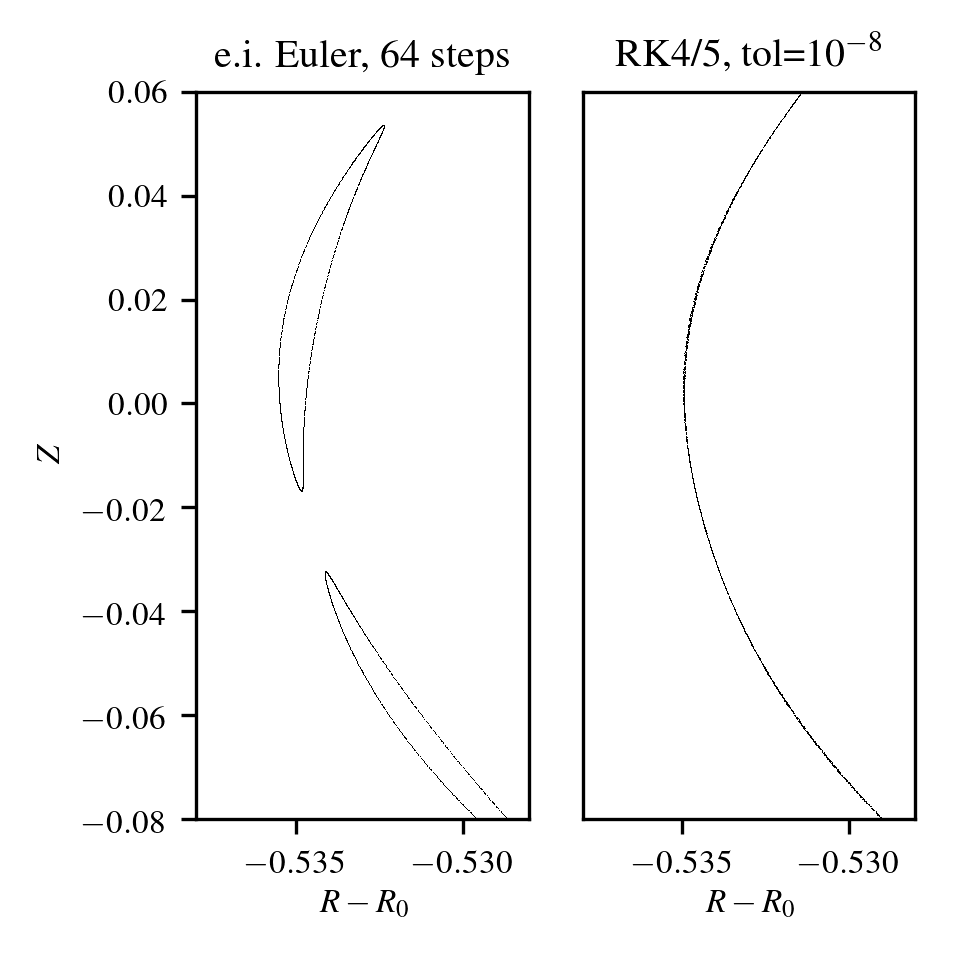}\includegraphics[scale=0.9]{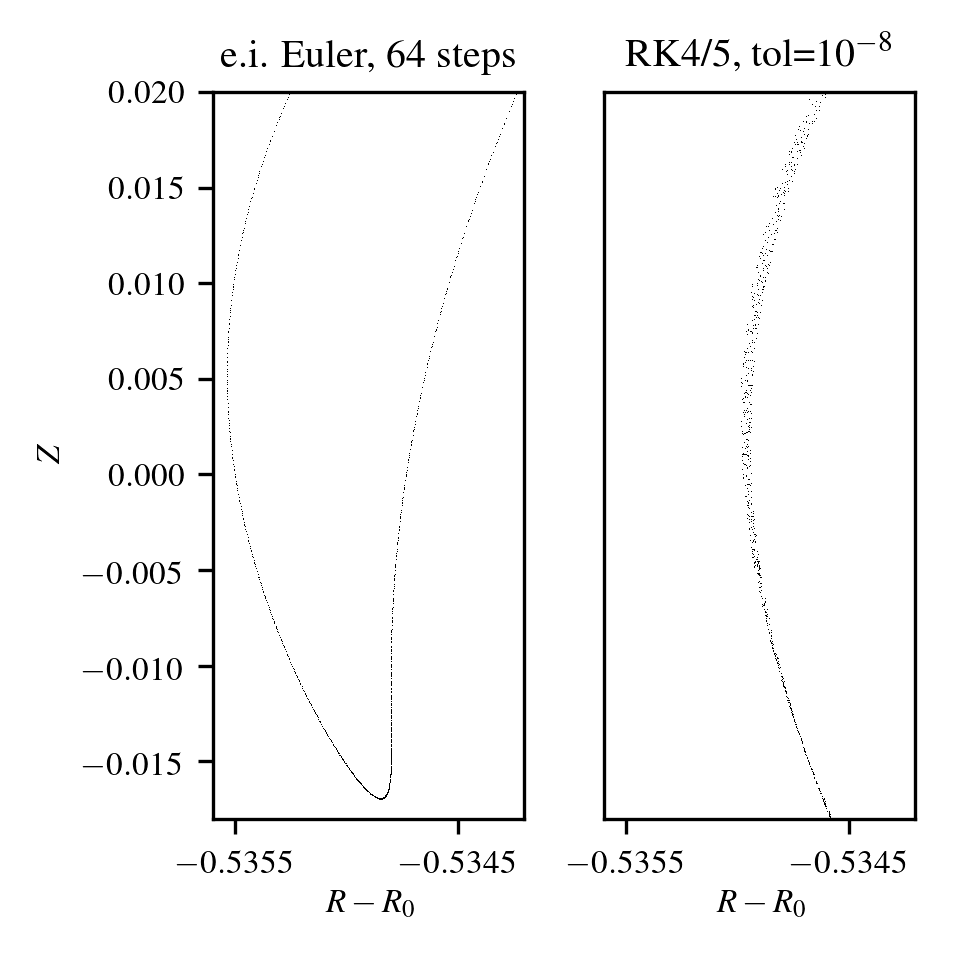}
\par\end{centering}
\caption{Poloidal projection of Poincaré sections at $v_{\parallel}=0$ switching
sign from $-$ to $+$. Upper left: full orbits, optically indistinguishable.
Remaining subplots: explicit-implicit Euler (left), RK4/5 (right).
Less accurate settings on the upper right show stochasticity of both
integrators. In the two zoom levels of the more accurate settings
on the lower plots one sees that the symplectic Euler integrator produces
a high-order island chain while RK still visibly scatters points on
this scale. \label{fig:Poloidal-projection}}
\end{figure}

For visualization of trapped orbits we introduce a Poincaré section
in phase-space which is a family of hypersurfaces containing turning
points defined by the condition $v_{\parallel}(\zset)=0$. Out of
two kinds of these surfaces we choose those where the sign of $v_{\parallel}$
changes from negative to positive. A return time to such a Poincaré
section defines the bounce period $\tau_{b}$. For the visual representation
we use a projection of footprints to the poloidal plane in quasi-cylindrical
coordinates $R$ and $Z$ defined via~(\ref{eq:cyl}) using the radial
variable $r=s$, where $s$ is the normalized toroidal magnetic flux.
In contrast to the axisymmetric model case those coordinates are strongly
distorted compared to their geometrical equivalent, but keep the topology
of usual cylindrical coordinates. Fig.~\ref{fig:Poloidal-projection}
shows results for a well trapped orbit started at $s=0.5$ and traced
until $t=1\,\mathrm{s}$ corresponding to $\approx80000$ bounce periods.
Computations from the explicit-implicit Euler scheme are compared
to an adaptive Runge-Kutta 4/5 (RK) scheme in the same canonicalized
flux coordinates at different accuracy settings. Namely the symplectic
Euler scheme is run at fixed time-steps with $32$ and $64$ steps
per magnetic field period estimated for strongly passing orbits moving
parallel to magnetic field lines. For the RK4/5 scheme relative tolerances
of $\mathrm{tol}=10^{-6}$ and $10^{-8}$ are used, respectively.
In the upper left of Fig.~\ref{fig:Poloidal-projection} footprints
of all four cases appear as a single curve. However, when zooming
closer, one can recognize a number of differences. In the less accurate
settings of 32 steps / $\mathrm{tol}=10^{-6}$ both integrators produce
scattered points already in the zoomed picture. The more accurate
symplectic setting of $64$ points produces a regular orbit consisting
of an island chain in the Poincaré section. \blue{This chain demonstrates
a kind of error made by the symplectic integrator at coarse resolution.
At half the step width a continuous line is produced instead, which
is not shown here.} RK4/5 results for tolerance $10^{-8}$ seem to
produce a perfectly regular orbit in the first zoom level. However,
when zooming closer, one can see that points are still scattered for
RK4/5, while islands of the symplectic Euler scheme appear as closed
lines.

\begin{figure}
\begin{centering}
\includegraphics[scale=0.9]{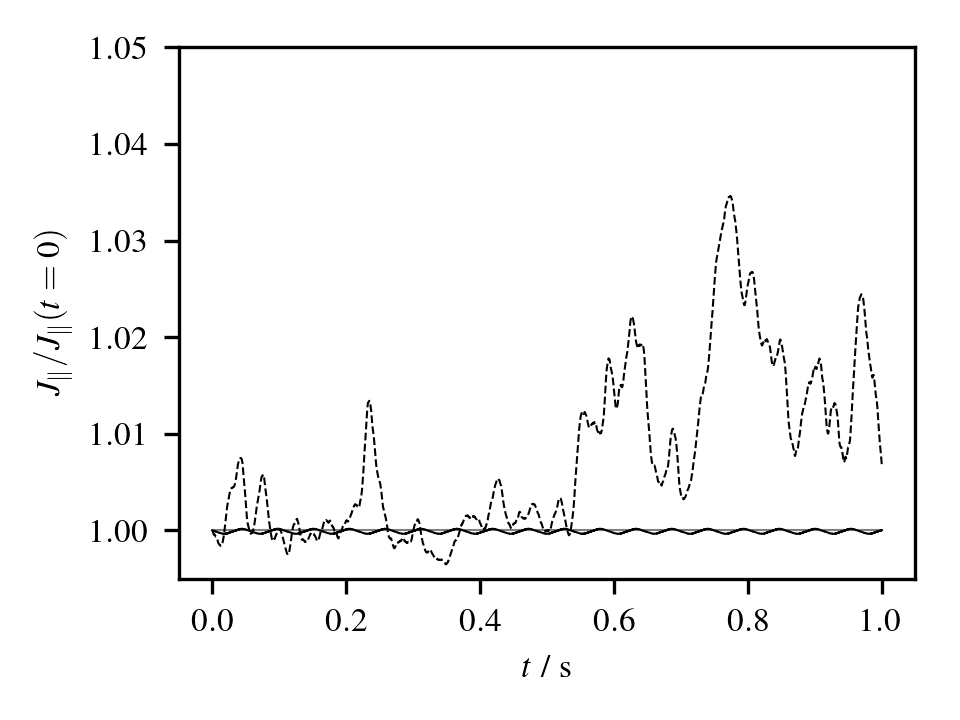}\includegraphics[scale=0.9]{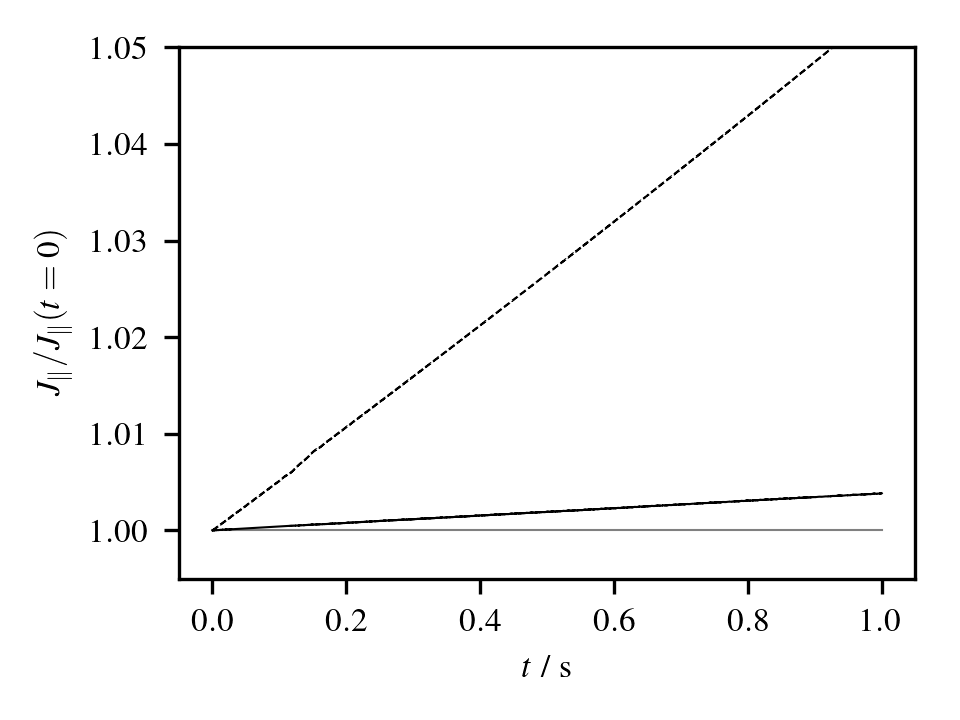}
\par\end{centering}
\caption{Parallel adiabatic invariant $J_{\parallel}$ of a fusion $\alpha$
orbit in a stellarator over physical time $t$. Here $J_{\parallel}$
is normalized to its initial value. Left: symplectic Euler 64/32 steps
(solid/dashed). Right: RK4/5 tol=$10^{-8}/10^{-6}$ (solid/dashed),
reference in gray. \label{fig:Parallel-adiabatic-invariant}}
\end{figure}

Fig.~\ref{fig:Parallel-adiabatic-invariant} shows the change over
time of the parallel adiabatic invariant $J_{\parallel}$ defined
in (\ref{eq:Jpar}), that should be well conserved for the exact motion
for the orbit used in the example. Conservation of this quantity is
of utmost importance for stellarator optimization~\citep{Mikhailov2002,Subbotin2006,Drevlak2014b}.
The Hamiltonian $H$ is not compared since it is exactly conserved
in the RK scheme due to the design of the used equations set \citep{Nemov2014},
and has only a minor influence on the orbit footprint. Here the difference
between the symplectic Euler scheme and the RK4/5 scheme becomes apparent.
In the more accurate setting of 64 steps per field period the symplectic
scheme conserves $J_{\parallel}$ up to small periodic oscillations.
In the less accurate setting of 32 steps, numerical \emph{diffusion
}of the symplectic integrator is visible, corresponding to chaotic
but still Hamiltonian motion. In contrast, the RK4/5 results \emph{drift}
away from the initial value, with smaller drift for the higher accuracy
setting, indicating a violation of Liouville's theorem by sources/sinks
in the phase flow present in non-symplectic schemes. In the latter
case of tolerance $10^{-8}$ the deviation of $J_{\parallel}$ after
the integration time of $1s$ is still relatively small, which explains
the satisfactory result for the orbit in Fig~\ref{fig:Poloidal-projection}.

\red{In contrast to the axisymmetric tokamak case, $J_{\parallel}$
is not an exact invariant in a stellarator. The minimum deviation
$\delta J_{\parallel}$ is limited by the 3D device geometry. Physically
correct behavior of numerical orbits can be expected when the numerical
error becomes lower than this threshold. This claim is supported by
the similar scaling of loss statistics in the subsequent section.
In Fig.~11 scaling laws of the conservation of $J_{\parallel}$ are
explored in a similar fashion to Fig.~\ref{fig:Jpartok}, but now
limited at a lower threshold value. Explicit-implicit Euler method
and single-stage midpoint rule are again favorable in terms of evaluations
$N_{\mathrm{f}}$ per bounce period with lowest CPU time for the Euler
scheme. As field evaluations become more expensive in 3D the Verlet
scheme now performs significantly worse than the single-stage midpoint
rule but still better than the original implicit midpoint rule of
Ref.~\citep{Zhang2014-32504}. Performance measures of Runge-Kutta
4/5 are well behind the ones of symplectic schemes.}

\blue{Similar plots of $\delta J_{\parallel}$ are given for fourth-order
methods in Fig.~\ref{fig:jparstell-1}. The compared methods are
the two-stage Gauss-Legendre scheme without evaluation at full steps,
the symmetric composition of first order schemes ``S, m=5'' of McLachlan~\citep{McLachlan1995}
pointed out in chapter V.3 of Ref.~\citep{Hairer2002-Geometric},
the three-stage Lobatto IIIA-IIIB pair, and again adaptive Runge-Kutta
4/5 as a comparison. There only the Gauss-Legendre method outperforms
RK4/5 but doesn't offer an advantage over lower-order explicit-implicit
Euler and single-stage midpoint rule in Fig.~\ref{fig:Jpartok} for
the purpose of conserving $J_{\parallel}$ in this setting.}

\begin{figure}
\begin{centering}
\includegraphics[width=0.48\textwidth]{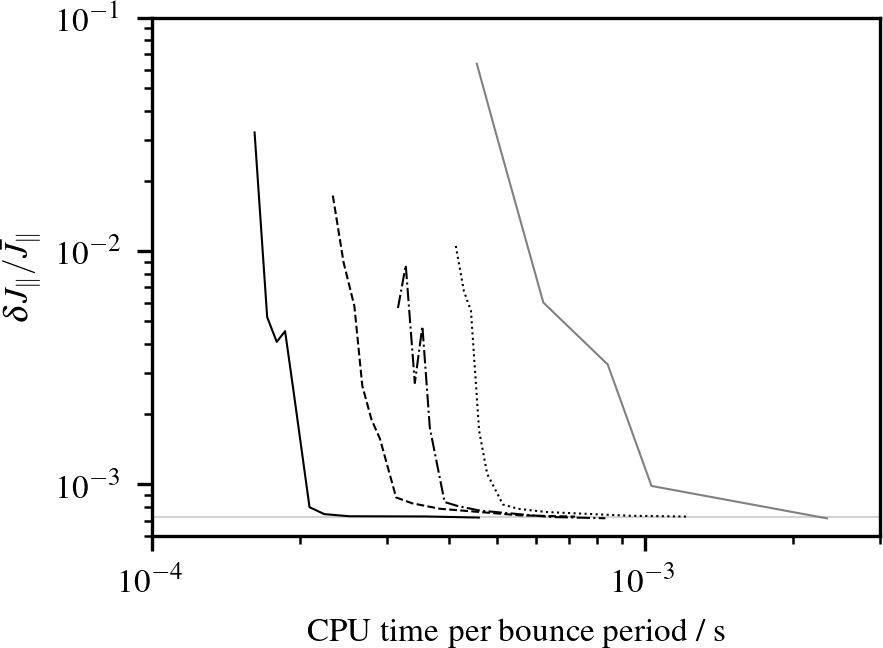}\quad{}\includegraphics[width=0.48\textwidth]{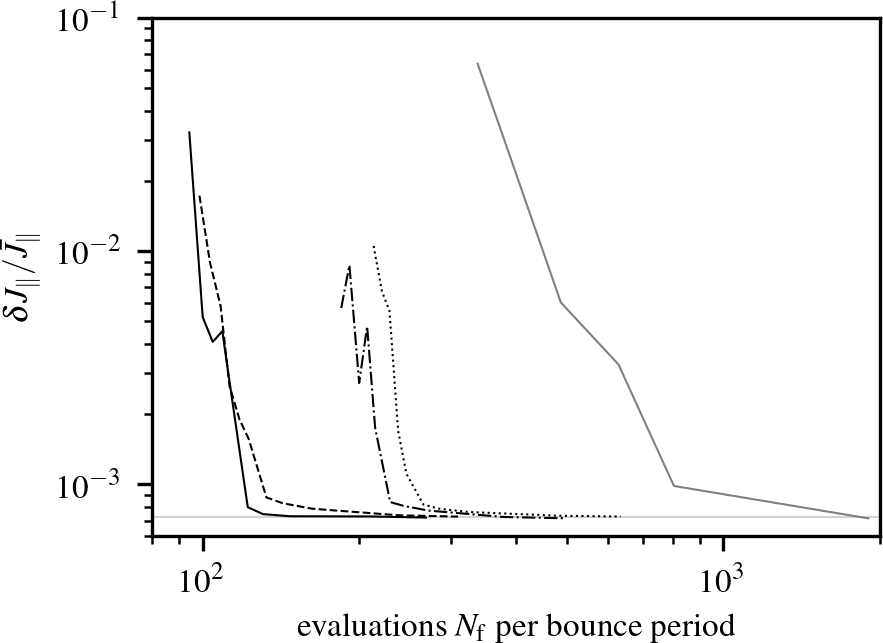}
\par\end{centering}
\caption{\red{Normalized root-mean-square deviation of the parallel invariant
$J_{\parallel}$ for a fusion $\alpha$ orbit in a 3D stellarator
field during $10^{5}$ bounce turns as in Fig.~\ref{fig:Jpartok}.
From left to right: explicit-implicit Euler, single-stage midpoint,
Verlet, midpoint, RK4/5. }\label{fig:jparstell}}
\end{figure}
\begin{figure}
\begin{centering}
\includegraphics[width=0.48\textwidth]{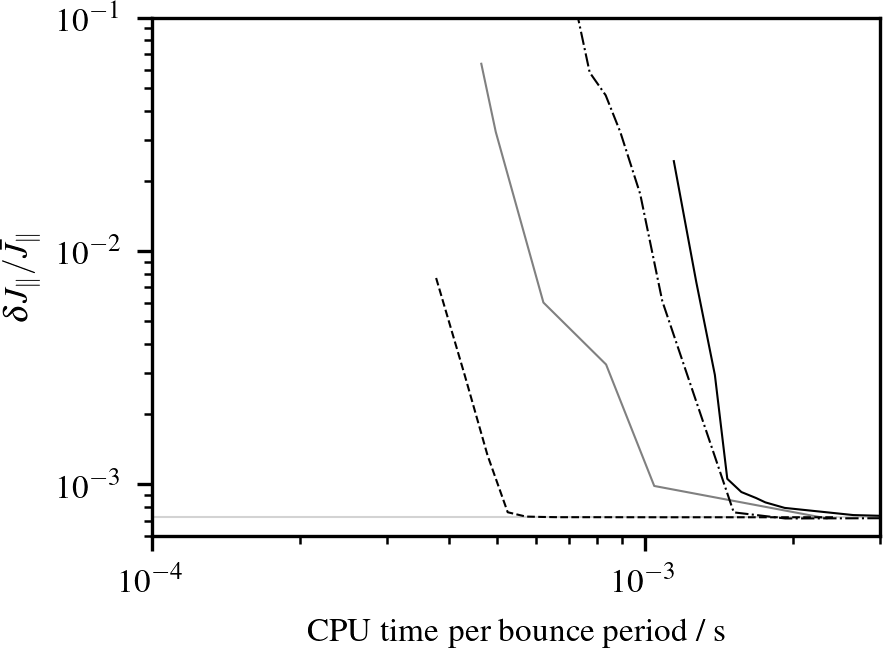}\quad{}\includegraphics[width=0.48\textwidth]{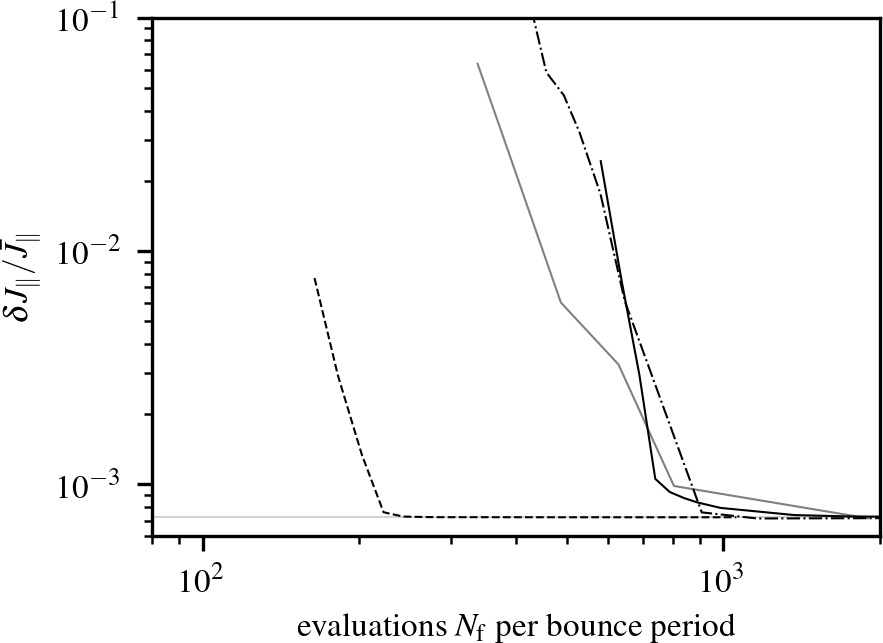}
\par\end{centering}
\caption{\red{Normalized root-mean-square deviation of the parallel invariant
$J_{\parallel}$ as in Fig.~\ref{fig:Jpartok} for fourth-order schemes.
Two-stage Gauss-Legendre (dashed), RK4/5 (solid gray), composition
method~\citep{McLachlan1995} (dash-dotted), Lobatto IIIA-IIB pair
(solid black). }\label{fig:jparstell-1}}
\end{figure}

\subsubsection{Loss statistics of fast fusion alpha particles}

Next, the loss fraction computed from a statistical ensemble of $1000$
fast fusion $\alpha$ particles after the physical time of $1s$ is
compared to a reference run with a Runge-Kutta 4/5 integrator in the
original (i.e. not canonicalized) magnetic flux coordinates at a relative
tolerance of $10^{-10}$. The $3.5\,\mathrm{MeV}$ fusion alpha particles
are initialized with uniform volumetric density on a magnetic surface
corresponding to normalized toroidal flux $s=0.75$ and with isotropic
distribution in velocity space. A particle is counted as lost if its
guiding-center orbit leaves the outer boundary of the device, which
is identified by a numerical check of geometrical bounds. Since orbits
are initialized here at one of the outer flux surfaces ($s=0.75$)
the resulting losses do not represent the quality of device optimization
but are used as a benchmark to compare statistics from different schemes.

\begin{figure}
\begin{centering}
\includegraphics[width=0.5\textwidth]{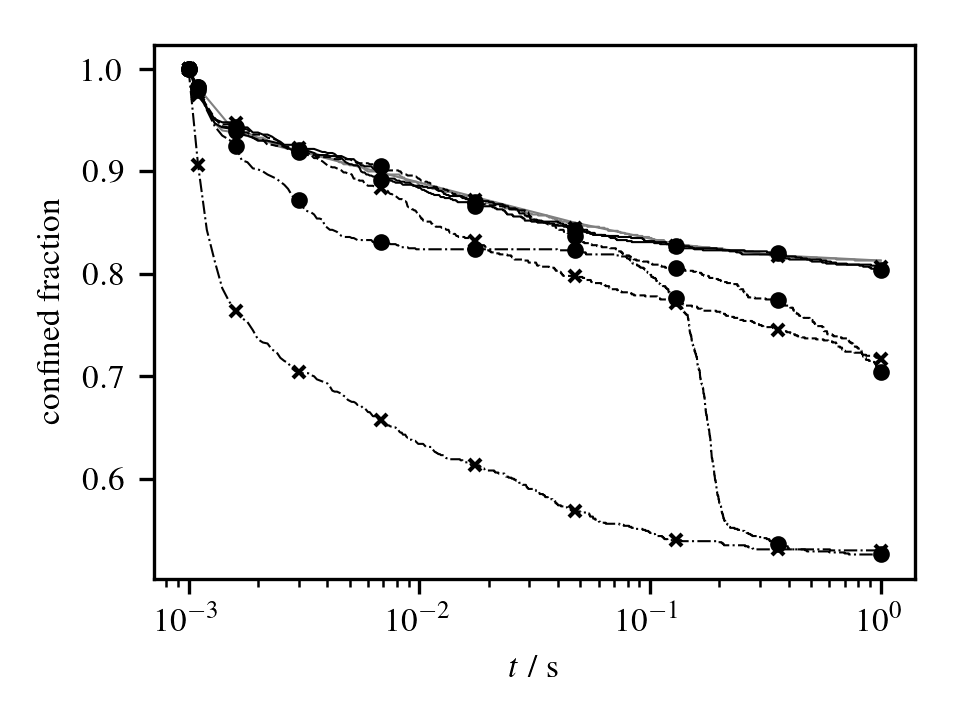}
\par\end{centering}
\caption{Loss statistics for a reference run (gray), RK4/5 in canonical flux
coordinates with tolerances $10^{-3}/10^{-6}/10^{-8}$ ($\boldsymbol{\mathbf{\times}}$,
dash-dotted/dashed/solid) and explicit-implicit Euler with $16/32/64$
steps per field period ($\bullet$, dash-dotted/dashed/solid). \label{fig:Loss-statistics-for}}
\end{figure}

Fig.~\ref{fig:Loss-statistics-for} shows the relative losses over
a logarithmic time scale. The importance of the conservation of $J_{\parallel}$
is directly reflected in those results, where a match to the reference
could be achieved by 64 steps per field period for symplectic Euler
and relative tolerance of $10^{-8}$ for RK4/5, leaving $J_{\parallel}$
conserved within bounds of less than one percent in Fig.~\ref{fig:Parallel-adiabatic-invariant}.
Final confined fractions in this settings at $t=1\,\mathrm{s}$ were
80.7\% (symplectic), 80.4\% (RK4/5) and 81.3\% (reference). At lower
accuracy the symplectic scheme loses orbits due to chaotic diffusion,
while in the RK4/5 scheme \sout{also} non-Hamiltonian mechanisms
\blue{also} play a role, leading to a more complicated time-dependency
of losses.

\begin{figure}
\begin{centering}
\includegraphics{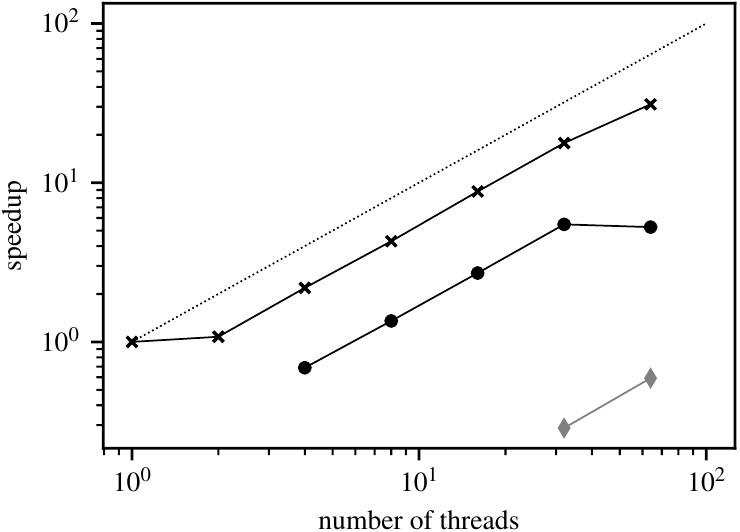}
\par\end{centering}
\caption{Scaling performance up to 64 threads on a DRACO node. Black: RK4/5
reference, tol=$10^{-10}$ ({\scriptsize{}$\blacklozenge$}, gray),
RK4/5 canonical coordinates, tol=$10^{-8}$ ($\bullet$), explicit-implicit
Euler, 64 steps per period ($\boldsymbol{\mathbf{\times}}$).\label{fig:Scaling-performance-on}}
\end{figure}

Finally we compare the performance and scaling of the different methods
for parallel computation at the described minimum accuracy reproducing
correct statistics (64 steps / tol=$10^{-8}$) in Fig.~\ref{fig:Scaling-performance-on}.
The numerical experiment has been performed on a single node of the
DRACO cluster of MPCDF with 32 CPU cores (Intel Xeon E5-2698) supporting
up to 64 concurrent threads with hyperthreading. While most of the
orbit loss algorithm can be parallelized in a trivial way over different
particles, shared memory access and other kinds of overhead can limit
the performance. Results indicate that the symplectic Euler scheme
suffers from parallelization overhead when increasing the number of
threads from 1 to 2, but scales linearly up to 64 cores, taking advantage
of hyperthreading. In contrast, the RK4/5 scheme in canonical flux
coordinates rather loses performance going from 32 to 64 threads.
This difference is possibly related to the less predictable behavior
of an adaptive RK scheme, making automatic compile-time and run-time
optimization more difficult. All in all this yields a speedup factor
of the symplectic Euler scheme of 3.2 (4-32 threads) and a maximum
of 5.9 (64 threads) compared to RK4/5 in canonical flux coordinates.
A speedup >50 is realized compared to the reference case. Runs on
less than respectively 4 and 32 threads were not performed for the
RK4/5 schemes due to limit wall-clock time for a single run to less
than 24h. The highest performance was achieved by the symplectic explicit-implicit
Euler scheme with 64 threads, reproducing correct loss statistics
on a single node within 10 minutes.

\begin{figure}
\begin{centering}
\includegraphics[width=0.5\textwidth]{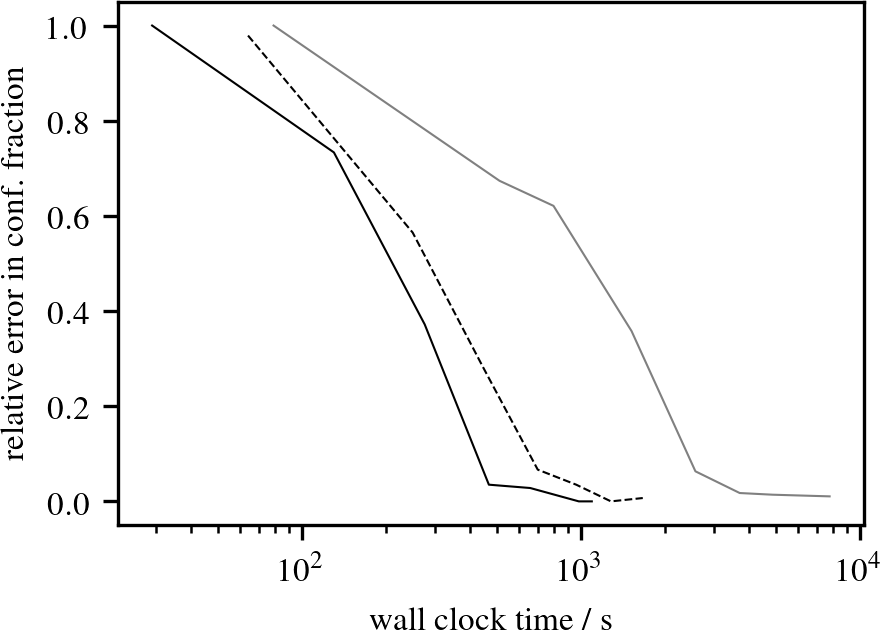}\includegraphics[width=0.5\textwidth]{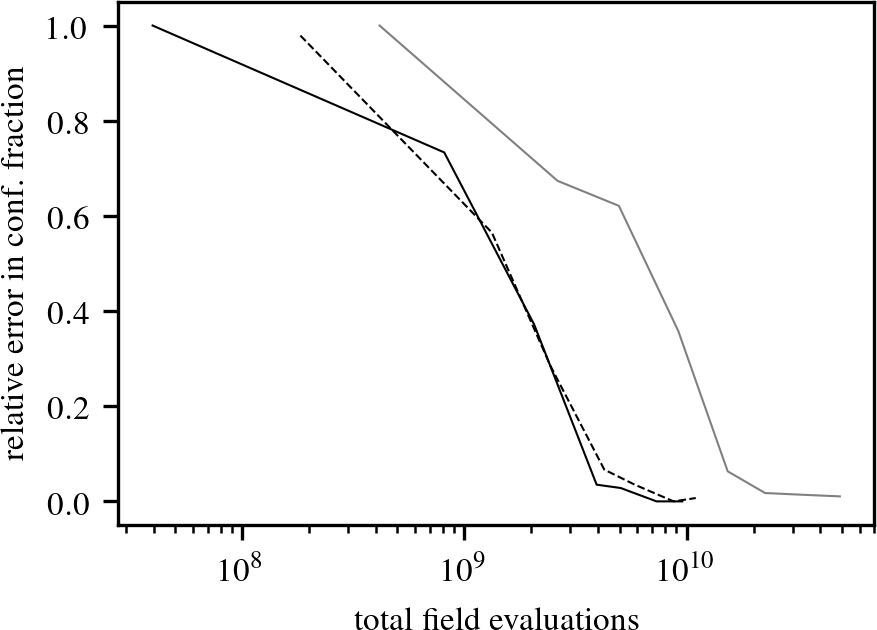}
\par\end{centering}
\caption{\red{Relative error in final confined fraction of $\alpha$ particles
after $1s$ physical time over computation wall clock time with 64
threads (left) and field evaluations (right), respectively. Expl.-impl.
Euler (solid black), single-stage midpoint (dashed), RK4/5 in canonical
coordinates (solid gray) at different resolutions.}\label{fig:Loss-statistics-for-1}}
\end{figure}

\red{The scaling of the relative error in the final confined $\alpha$
fraction in terms of wall-clock time and total number of field evaluations
is shown in Fig.~\ref{fig:Loss-statistics-for-1} for the maximum
setting of $64$ threads. There, in addition to explicit-implicit
Euler and RK4/5 schemes, also the single-stage midpoint rule is compared
in terms of performance. The results essentially reflect the behavior
of the parallel invariant in Fig.~\ref{fig:jparstell}. While the
number of field evaluations to reach a certain accuracy is roughly
the same for symplectic Euler and single-stage midpoint schemes, the
simpler Euler method is faster in terms of CPU time by a factor of
about $1.5$ in the presented application case. Both symplectic methods
outperform the non-symplectic RK4/5 scheme.}

\newpage{}

\section{Summary \red{and outlook}\label{sec:Summary}}

The construction of symplectic numerical integration methods with
\green{only} internal stages in non-canonical coordinates has been
presented and applied to guiding-center motion in 2D and 3D magnetic
field configuration with nested flux surfaces. The method relies on
a existing symplectic schemes in canonical coordinates and finding
the pertinent non-canonical representation of quadrature points in
phase-space by solving implicit equations. \green{Originally fully
implicit methods of Ref.~\citep{Zhang2014-32504} have been generalized
to arbitrary order partitioned Runge-Kutta schemes together with optimizations
to enhance their performance.} \red{In addition an optimized single-stage
variant of the implicit midpoint rule has been presented that reduces
the size of the implicit system by half.} \blue{Depending on the
problem at hand, the presented methods could offer an advantage compared
to projected variational integrators of Ref.~\citep{Kraus2017-preprint}
that have to solve a larger set of implicit equations to satisfy projection
constraints. Degenerate variational integrators of Ref.~\citep{Ellison2018}
don't require these additional equations, but only first-order schemes
have been constructed so far.}

A suitable transformation to canonical coordinates for the guiding-center
Lagrangian in 3D magnetic fields given in straight field line flux
coordinates has been provided \red{for time-independent magnetic
geometry}. Since it is possible to numerically evaluate this purely
spatial transformation in a \red{parallelized} preprocessing step
the added computational overhead is negligible for long-term tracing
of orbits in a stationary equilibrium field. In particular this is
the case for the computation of fast fusion $\alpha$ particle losses
in optimized stellarators with magnetic configurations computed by
an equilibrium code such as VMEC~\citep{Hirshman1983}. \blue{For
future studies a regularized guiding-center Lagrangian~\citep{Burby2017-110703}
could be an interesting alternative to construct a transformation
to canonical coordinates also for equilibria without nested flux surfaces.}

Long-term conservation of energy and parallel adiabatic invariant
has been demonstrated as an intrinsic feature of symplectic schemes,
where the latter property is of high importance for stellarator optimization.
As a simple model we first considered an analytical 2D tokamak field,
demonstrating similar to superior performance of different symplectic
methods compared to the commonly used Runge-Kutta 4/5 scheme. As a
real-world application case a realistic 3D field of an optimized stellarator
has been considered. \red{In this setting symplectic Euler and \sout{Störmer-}Verlet
schemes have been shown to be competitive to the implicit midpoint
rule. The simple explicit-implicit Euler scheme could reach the best
performance in terms of conservation of the parallel invariant. For
that purpose fourth order partitioned methods did not show any advantage
over the Gauss-Legendre method or lower order methods. They could
however be useful to reduce the implicit problem size in timesteps
of systems with more degrees of freedom, e.g. interacting many-particle
systems.}

Finally the loss fraction of fusion $\alpha$ particles \red{from
an optimized stellarator configuration} has been computed based on
a sufficiently large ensemble of 1000 orbits with 480 trapped orbits
traced over their slowing down time of one second. \red{The link
of sufficiently accurate conservation of the parallel adiabatic invariant
by a numerical integration scheme to final statistical accuracy has
been illustrated.} \sout{These} \red{Fusion $\alpha$ particle}
losses are well reproduced by the first order explicit-implicit Euler
method compared to a highly accurate reference computation. On 64
threads with shared memory access on a single computing node the explicit-implicit
Euler scheme could achieve this task within a wall-clock time of $\approx10$
minutes, compared to \red{$\approx15$ minutes for the single-stage
midpoint rule and} $\approx1\,\mathrm{h}$ for a Runge-Kutta scheme
with the same statistical accuracy. This puts the time to compute
fast fusion $\alpha$ particle losses in a range useful for stellarator
optimization where the computation time for magnetic equilibria is
of the same order.
\newpage
\section*{Acknowledgments}

The authors would like to thank Benedikt Brantner for supporting initial
investigations, Michael Drevlak for providing stellarator field configurations
and John Cary, Martin Heyn, Michael Kraus, Stefan Possanner and Don
Spong for useful discussions. Support from NAWI Graz, from the OeAD
under the WTZ grant agreement with Ukraine No UA 04/2017 and funding
from the Helmholtz Association within the research project ``Reduced
Complexity Models'' is gratefully acknowledged. This work has been
carried out within the framework of the EUROfusion Consortium and
has received funding from the Euratom research and training programme
2014-2018 and 2019-2020 under grant agreement No 633053. The views
and opinions expressed herein do not necessarily reflect those of
the European Commission.

\section*{}

\bibliographystyle{elsarticle-num}
\bibliography{paper_sympl}

\appendix

\section{Evaluation of non-canonical variables at full timesteps~\label{subsec:Evaluation-of-non-canonical}}

For post-processing integration results it is sometimes needed to
compute non-canonical variables at full time-steps,
\begin{align}
\zset_{(n)} & \equiv\zset(\qset_{(n)},\pset_{(n)}),\quad\zset_{(n+1)}\equiv\zset(\qset_{(n+1)},\pset_{(n+1)})
\end{align}
from intermediate values used in the symplectic Euler variants. As
mentioned those equations are only defined implicitly, but can be
approximated by a Taylor expansion at sufficiently small distances
in phase-space. For explicit-implicit Euler with
\begin{align}
\zset_{(n-1,\ei)} & =\zset(\qset_{(n-1)},\pset_{(n)}),\quad\zset_{(n,\ei)}=\zset(\qset_{(n)},\pset_{(n+1)})
\end{align}
one can use a forward or backward expansion,
\begin{align}
z_{(n+1)}^{\alpha} & \approx z_{(n,\ei)}^{\alpha}+(q_{(n+1)}^{i}-q_{(n)}^{i})\frac{\partial z_{(n,\ei)}^{\alpha}}{\partial q^{i}},\label{eq:rfull}\\
z_{(n)}^{\alpha} & \approx z_{(n,\ei)}^{\alpha}-(p_{j,(n+1)}-p_{j,(n)})\frac{\partial z_{(n,\ei)}^{\alpha}}{\partial p_{j}},
\end{align}
or a second-order centered expansion
\begin{align}
z_{(n)}^{\alpha} & \approx\frac{z_{(n-1,\ei)}^{\alpha}+z_{(n,\ei)}^{\alpha}}{2}+\frac{q_{(n)}^{i}-q_{(n-1)}^{i}}{2}\frac{\partial z_{(n-1,\ei)}^{\alpha}}{\partial q^{i}}-\frac{p_{j,(n+1)}-p_{j,(n)}}{2}\frac{\partial z_{(n,\ei)}^{\alpha}}{\partial p_{j}}.
\end{align}
Derivatives are as usually obtained from the inverse Jacobian of the
transformation from $\zset$ to $\qset,\pset$. For implicit-explicit
Euler with
\begin{align}
\zset_{(n-1,\ei)} & \equiv\zset(\qset_{(n)},\pset_{(n-1)}),\quad\zset_{(n,\ei)}\equiv\zset(\qset_{(n+1)},\pset_{(n)})
\end{align}
we obtain instead
\begin{align}
z_{(n+1)}^{\alpha} & \approx z_{(n,\ei)}^{\alpha}+(p_{j,(n+1)}-p_{j,(n)})\frac{\partial z_{(n,\ei)}^{\alpha}}{\partial p_{j}},\label{eq:rfull-1}\\
z_{(n)}^{\alpha} & \approx z_{(n,\ei)}^{\alpha}-(q_{(n+1)}^{i}-q_{(n)}^{i})\frac{\partial z_{(n,\ei)}^{\alpha}}{\partial q^{i}},
\end{align}
and
\begin{align}
z_{(n)}^{\alpha} & \approx\frac{z_{(n-1,\ei)}^{\alpha}+z_{(n,\ei)}^{\alpha}}{2}+\frac{p_{j,(n+1)}-p_{j,(n)}}{2}\frac{\partial z_{(n,\ei)}^{\alpha}}{\partial p_{j}}-\frac{q_{(n)}^{i}-q_{(n-1)}^{i}}{2}\frac{\partial z_{(n-1,\ei)}^{\alpha}}{\partial q^{i}}.
\end{align}
When approximating full-step results for the second order \sout{Störmer-}Verlet/leap-frog
scheme it is convenient to use half-step quantities on both sides
of a full timestep and otherwise proceed in the same way as for the
partitioned Euler schemes.

\section{Second derivatives for guiding-center motion~\label{subsec:Jacobian-for-root-finding}}

Here second derivatives of parallel velocity $v_{\parallel}$, poloidal
canonical momentum $p_{\tht}$ and Hamiltonian $H$ with respect to
$\zset=(r,\tht,\ph,p_{\ph})=(\xset,p_{\ph})$ are provided. Those
are required in order to solve guiding-center equations such as (\ref{eq:FEuler}-\ref{eq:F2Euler})
and (\ref{eq:FEuler-1}-\ref{eq:F2Euler-1}) implicitly by schemes
relying on user-supplied second derivatives. For $v_{\parallel}$
terms containing partial derivatives $\partial/\partial p_{\ph}$
with respect to $p_{\ph}$ are
\begin{align}
\frac{\partial^{2}v_{\parallel}}{\partial p_{\ph}\partial x^{i}}= & -\frac{1}{mh_{\ph}^{\,2}}\frac{\partial h_{\ph}}{\partial x^{i}},\qquad\frac{\partial^{2}v_{\parallel}}{\partial p_{\ph}^{\,2}}=0.
\end{align}
For second spatial derivatives we use
\begin{align}
\frac{\partial^{2}}{\partial x^{i}\partial x^{j}}(h_{\ph}v_{\parallel}) & =-\frac{e}{mc}\frac{\partial^{2}A_{\ph}}{\partial x^{i}\partial x^{j}},
\end{align}
so
\begin{equation}
\frac{\partial^{2}v_{\parallel}}{\partial x^{i}\partial x^{j}}=-\frac{1}{h_{\ph}}\left(\frac{e}{mc}\frac{\partial^{2}A_{\ph}}{\partial x^{i}\partial x^{j}}+\frac{\partial^{2}h_{\ph}}{\partial x^{i}\partial x^{j}}v_{\parallel}+\frac{\partial h_{\ph}}{\partial x^{i}}\frac{\partial v_{\parallel}}{\partial x^{j}}+\frac{\partial h_{\ph}}{\partial x^{j}}\frac{\partial v_{\parallel}}{\partial x^{i}}\right).
\end{equation}
For $p_{\tht}$ second derivatives involving $\partial/\partial p_{\ph}$
are
\begin{align}
\frac{\partial^{2}p_{\tht}}{\partial x^{i}\partial p_{\ph}}=\frac{1}{h_{\ph}}\frac{\partial h_{\tht}}{\partial x^{i}}-\frac{h_{\tht}}{h_{\ph}^{\,2}}\frac{\partial h_{\ph}}{\partial x^{i}} & ,\qquad\frac{\partial^{2}p_{\tht}}{\partial p_{\ph}^{\,2}}=0.
\end{align}
Second spatial derivatives of $p_{\tht}$ follow as
\begin{equation}
\frac{\partial^{2}p_{\tht}}{\partial x^{i}\partial x^{j}}=m\left(\frac{\partial^{2}v_{\parallel}}{\partial x^{i}\partial x^{j}}h_{\tht}+\frac{\partial v_{\parallel}}{\partial x^{i}}\frac{\partial h_{\tht}}{\partial x^{j}}+\frac{\partial v_{\parallel}}{\partial x^{j}}\frac{\partial h_{\tht}}{\partial x^{i}}+v_{\parallel}\frac{\partial^{2}h_{\tht}}{\partial x^{i}\partial x^{j}}\right)+\frac{e}{c}\frac{\partial^{2}A_{\tht}}{\partial x^{i}\partial x^{j}}.
\end{equation}
For $H$ second derivatives involving $\partial/\partial p_{\ph}$
are
\begin{align*}
\frac{\partial^{2}H}{\partial p_{\ph}\partial x^{i}} & =m\left(\frac{1}{h_{\ph}}\frac{\partial v_{\parallel}}{\partial x^{i}}-\frac{v_{\parallel}}{h_{\ph}^{\,2}}\frac{\partial h_{\ph}}{\partial x^{i}}\right),\qquad\frac{\partial^{2}H}{\partial p_{\ph}^{\,2}}=0.
\end{align*}
Second spatial derivatives of $H$ are
\begin{equation}
\frac{\partial H^{2}}{\partial x^{i}\partial x^{j}}=m\left(v_{\parallel}\frac{\partial^{2}v_{\parallel}}{\partial x^{i}\partial x^{j}}+\frac{\partial v_{\parallel}}{\partial x^{i}}\frac{\partial v_{\parallel}}{\partial x^{j}}\right)+\mu\frac{\partial B}{\partial x^{i}\partial x^{j}}+e\frac{\partial\Phi}{\partial x^{i}\partial x^{j}}.
\end{equation}

~
\end{document}